\documentclass[aps, prd, superscriptaddress, nofootinbib, 10pt]{revtex4-2}

\usepackage[english]{babel}
\usepackage[T1]{fontenc}

\usepackage{emptypage} 
\usepackage{graphicx} 

\usepackage{amsmath}
\usepackage{comment}

\usepackage[caption=false]{subfig}

\usepackage{float} 
\usepackage[dvipsnames]{xcolor} 
\usepackage[edges]{forest}
\usepackage{nicematrix} 
\usepackage[compat = 1.1.0]{tikz-feynman} 

\usepackage[a-3b]{pdfx} 

\allowdisplaybreaks

\NiceMatrixOptions{cell-space-limits = 7pt}

\newcommand*{\Ohat}{\skew{2}{\widehat}{O}}
\newcommand*{\phat}{\skew{2}{\hat}{p}}
\newcommand*{\epsilonhat}{\skew{2}{\hat}{\epsilon}}
\newcommand*{\psitilde}{\skew{2}{\tilde}{\psi}}
\newcommand*{\Ntilde}{\skew{2}{\tilde}{N}}
\newcommand*{\betatilde}{\skew{2}{\tilde}{\beta}}
\newcommand*{\gammatilde}{\skew{2}{\tilde}{\gamma}}

\begin{document}

\title{A Light-Front Model for the Transition Distribution Amplitudes for Backward Timelike Compton Scattering}

\author{Barbara Pasquini}
\thanks{Electronic address: barbara.pasquini@unipv.it}
\affiliation{Dipartimento di Fisica, Università degli Studi di Pavia, I-27100 Pavia, Italy} 
\affiliation{Istituto Nazionale di Fisica Nucleare, Sezione di Pavia, I-27100 Pavia, Italy}

\author{Andrea Schiavi\footnote{Currently employed in high-school teaching.}}
\thanks{Electronic address: andrearschiavi@gmail.com}
\affiliation{Dipartimento di Fisica, Università degli Studi di Pavia, I-27100 Pavia, Italy}

\date{\today}


\begin{abstract}

\noindent To access information on the internal structure of the nucleon, data from a variety of scattering experiments can be analyzed, in regimes where the information factorizes from an otherwise known scattering amplitude. A recent development, promising new insight, is the study of exclusive reactions in the backward kinematical region, where the information can be encoded in Transition Distribution Amplitudes (TDAs). We model the photon-to-nucleon TDAs, entering the factorized description of backward Timelike Compton Scattering, using techniques of light-front dynamics to integrate information from a quark model for the photon and the nucleon. We include the results of numerical predictions that could inform further experiments at Jefferson Lab and the future Electron--Ion Collider.

\end{abstract}

\maketitle

\section{Introduction}

Hard exclusive processes offer invaluable insight into unraveling the parton structure of hadrons. Notable examples include Deeply-Virtual Compton Scattering (DVCS) and its time-reversal conjugate process, Timelike Compton Scattering (TCS). The former is the scattering of a high-virtuality spacelike photon off a nucleon target, resulting in the production of a real photon and the recoiling nucleon, while the latter  sees a real photon scatter off the nucleon target into a high-virtuality timelike photon. In the forward kinematical region, characterized by small absolute values of the Mandelstam variable $t$ and large absolute values of the Mandelstam variable $u$, information on the internal structure of the nucleon is encoded in Generalized Parton Distributions 
(GPDs)~\cite{Muller:1994ses, Ji:1996nm, Radyushkin:1997ki, Goeke:2001tz, Diehl_2003, Belitsky:2005qn, Boffi:2007yc,Berger:2001xd, Pire:2011st, Mueller:2012sma, Moutarde:2013qs}. These are related to matrix elements of a bilocal operator between the initial and final nucleon states, and represent the amplitude of transferring momentum to the hadron through the exchange of two partons.

The situation is more complex in the backward kinematical region, where $ \lvert u \rvert $ is small and $ \lvert t \rvert $ is large. However, the analogy with forward DVCS and TCS suggests a description in terms of a soft amplitude factorized from the hard scattering of the partons with the probe. The variable $u$ characterizes the transition between a real photon and a nucleon, which can be encoded in Transition Distribution Amplitudes (TDAs)~\cite{Pire:2021hbl, Pire:2022fbi, Pire:2022kwu}. These are related to matrix elements of a trilocal operator between the photon and nucleon states, and represent the amplitude of transferring momentum and one unit of baryon number through the exchange of three partons.

The focus of the present work is on backward TCS, which remains relatively unexplored compared to other processes involving 
TDAs~\cite{Lansberg:2011aa, Lansberg:2012ha, Pire:2013jva, Pire:2015kxa, Pasquini:2009ki}. The experimental study of TCS is a recent development, with data published for the first time in 2021 by the CLAS collaboration at Jefferson Lab for the forward 
region~\cite{CLAS:2021lky}. Moreover, backward TCS is expecially appealing, since the electromagnetic Bethe--Heitler background, where the initial photon directly couples to a lepton--antilepton pair in the final state, is significantly suppressed (except for very narrow regions of solid angle for the produced lepton)~\cite{Pire:2022fbi}. To compare the factorized description against experimental results and to guide further phenomenological studies, a model for the photon-to-nucleon TDAs is required, beginning with the leading contributions.

The model developed in this work is based on the framework of light-front dynamics (LFD)~\cite{Dirac:1949cp, Brodsky:1989pv, Brodsky:1997de, Kovchegov:2012mbw}, where the interacting particles are described in the Fock space in terms of light-front wave functions (LFWFs). In Sec.~\ref{section_btcs}, the backward kinematical region of TCS is analyzed, and the factorized description of the scattering amplitude is introduced. The photon-to-nucleon TDAs are defined, and their expressions in terms of matrix elements of a trilocal operator between the initial photon and the final nucleon states are derived. Section~\ref{section_model_tdas} is dedicated to modelling the photon-to-nucleon TDAs, specifically in the support region where the description in terms of the leading Fock-components of the photon and nucleon LFWFs is suitable. The photon is treated as a light quark--antiquark pair, while the Fock representation of the nucleon is truncated to the three valence quarks in a constituent quark model that has already been applied to GPDs~\cite{Boffi:2002yy, Boffi:2003yj, Pasquini:2004gc, Pasquini:2005dk, Pasquini:2006dv, Pasquini:2007xz} and to nucleon-to-neutral-pion 
TDAs~\cite{Pasquini:2009ki}. Given the impracticality of the complete formulas, the analytical results of the model calculation are illustrated schematically, giving the various components alongside instructions for combining them into the final formulas. Numerical predictions for a selected set of TDAs and their first Mellin moments are discussed in Sec.~\ref{subsec_numerical_results}. Concluding remarks and an outlook on further developments are given in the final Section~\ref{section_conclusions}.

\section{Backward Timelike Compton Scattering}
\label{section_btcs}

\subsection{The Backward Kinematical Region}

Timelike Compton Scattering refers to the scattering of a photon ($ \gamma $) off a nucleon ($N$) into a virtual photon $ \left( \gamma^{*} \right)$ and the recoiling nucleon $ \left( N' \right) $, schematically
\begin{equation}
\gamma \! \left( q \right) + N \! \left( p_{N} \right) \rightarrow \gamma^{*} \! \left( q' \right) + N' \! \left( p_{N}' \right) \! ,
\label{tcs}
\end{equation}
where in brackets are the four-momenta of the particles. The virtual photon is timelike and produces a lepton-antilepton pair in the final state. The Mandelstam variables of the process are defined as
\begin{alignat}{2}
s & = \left( q + p_{N} \right)^{2} & & = \left( q' + p'_{N} \right)^{2} \! ,
\label{mandelstams_tcs} \\
t & = \left( p_{N}' - p_{N} \right)^{2} & & = \left( q' - q \right)^{2} \! ,
\label{mandelstamt_tcs} \\
u & = \left( p_{N}' - q \right)^{2} & & = \left( q' - p_{N} \right)^{2} \! .
\label{mandelstamu_tcs}
\end{alignat}
We are interested in the kinematical region where
\begin{equation}
\left( q' \right)^{2} \! , s, \lvert t \rvert \gg \lvert u \rvert, m_{N}^{2},
\label{near_backward_region}
\end{equation}
where $m_{N}$ is the mass of the nucleon.

Introducing light-cone (LC) coordinates, for an arbitrary four-vector $v$ we write $v^{ \pm } = \left( v^{0} \pm v^{3} \right)$ and $\vec{v}_{ \perp} = \left( v^{1}, v^{2} \right)$, and give all components as $v^{ \mu } = \left( v^{+}, v^{-}, \vec{v}_{ \perp } \right)$. Furthermore, we adopt the $T$ subscript for the transverse part of the four-vector, i.e., $v_{T} = \left( 0, 0, \vec{v}_{ \perp } \right)$. The LC decomposition of a four-vector can be written in a Lorentz covariant fashion using two light-like vectors $p = \left( 1, 0, \vec{0}_{ \perp } \right)$ and $n = \left( 0, 1, \vec{0}_{ \perp } \right)$. We have 
\begin{equation}
v^{ \mu } = v^{+}p^{ \mu } + v^{-}n^{ \mu } + v_{T}^{ \mu},
\label{sudakov_decompo}
\end{equation}
where $v^{+} = 2 \left( n \cdot v \right) $, $v^{-} = 2 \left( p \cdot v \right) $, and $p \cdot v_{T} = n \cdot v_{T} = 0$.

In a reference frame where the $z$-axis is along the direction of the colliding real photon and proton, the momenta involved in the process are
\begin{alignat}{1}
q & = \left( 1 + \xi \right) \! p,
\label{sudakov_q} \\
p_{N} & = \frac{m_{N}^{2} \! \left( 1 + \xi \right)}{s - m_{N}^{2}} p + \frac{s - m_{N}^{2}}{\left( 1 + \xi \right)} n,
\label{sudakov_pn} \\
p'_{N} & = \left( 1 - \xi \right) \! p + \frac{m_{N}^{2} - \Delta_{T}^{2}}{\left( 1 - \xi \right)} n + \Delta_{T},
\label{sudakov_pnprime} \\
\Delta = p'_{N} - q & = -2 \xi p + \frac{m_{N}^{2} - \Delta_{T}^{2} }{ \left( 1 - \xi \right) } n + \Delta_{T},
\label{sudakov_delta} \\
q' & = p_{N} - \Delta,
\label{sudakov_qprime}
\end{alignat}
where $\xi$ is the skewness variable, with $0 \le \xi < 1$. It will also be useful to define the mean momentum
\begin{equation}
P = \frac{p'_{N} + q}{2} = p + \frac{m_{N}^{2} - \Delta_{T}^{2}}{2 \! \left( 1 - \xi \right) } n + \frac{ \Delta_{T} }{2},
\label{sudakov_mean}
\end{equation}
so that
\begin{equation}
p^{+} = P^{+},
\label{longit_sudakovp_and_meanmom}
\end{equation}
and
\begin{equation}
\xi = - \frac{ \left( p'_{N} - q \right) \cdot n}{ \left( p'_{N} + q \right) \cdot n} = - \frac{ \Delta \cdot n}{2P \cdot n}.
\label{xi}
\end{equation}
We also have 
\begin{equation}
u = \Delta^{2} = -2 \xi \frac{m_{N}^{2} - \Delta_{T}^{2} }{ \left( 1 - \xi \right) } + \Delta_{T}^{2},
\label{sudakov_u}
\end{equation}
so $ \lvert u \rvert $ is smaller for $ \Delta_{T} = 0$, i.e., when the trajectories of the outgoing particles are aligned with the incoming particles, and for $ \xi $ close to zero.

\subsection{Interpretation of the Photon-to-Nucleon Transition Distribution Amplitudes}

\begin{figure}[h]
\centering
\includegraphics[width = 0.5\textwidth, keepaspectratio]{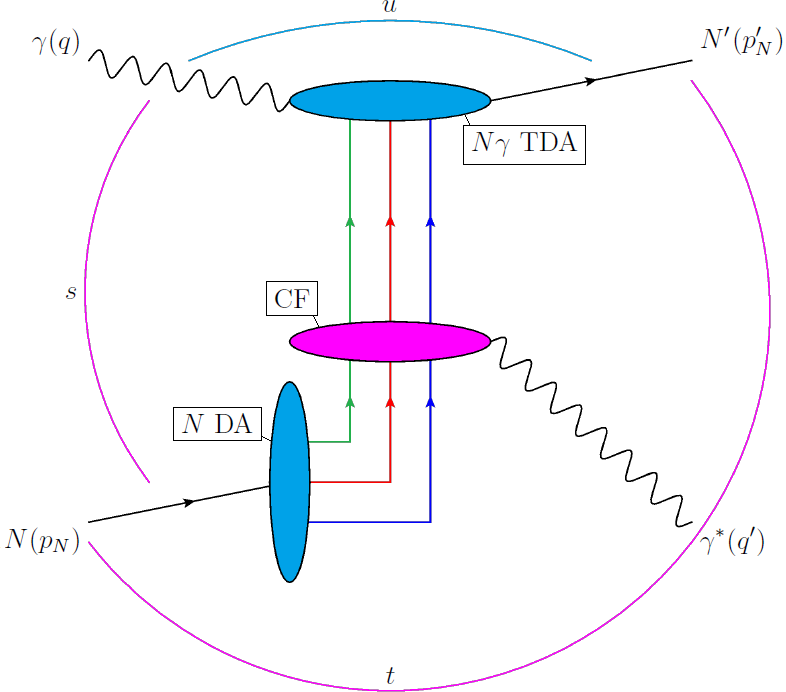}
\caption{(Color online.) Sketch of the factorized amplitude of backward TCS. The central (pink) oval shape represents the high-energy elementary scattering, described by CFs. The upper and lower (light blue) oval shapes represent the non-perturbative subprocesses, encoded in photon-to-nucleon TDAs and nucleon DAs, respectively.}
\label{btcs_fig}
\end{figure}

Following Refs.~\cite{Pire:2021hbl, Pire:2022fbi}, we apply a collinear factorized description of the backward TCS in the  kinematical region~\eqref{near_backward_region}. Owing to the high-energy scale of the final photon and the low-energy scale $ \lvert u \rvert = \lvert p_{N}' - q \rvert^{2}$, we can isolate from the rest of the scattering amplitude a transition from the initial photon to the final nucleon. We can imagine this transition happening through an exchange of partons between the initial photon and the initial nucleon, which, once extracted, take part in a perturbative subprocess that produces the final photon. This elementary scattering is  described by coefficient functions (CFs). The splitting of the initial nucleon into its constituent quarks is described by  non-perturbative objects called nucleon Distribution Amplitudes (DAs), while the transition from the initial photon to the final nucleon is encoded in photon-to-nucleon ($N \gamma $) TDAs. This factorized description of the amplitude of backward TCS is sketched in Fig.~\ref{btcs_fig}.

\begin{figure}[h]
\centering
\subfloat[]{\includegraphics[width = 0.32\textwidth]{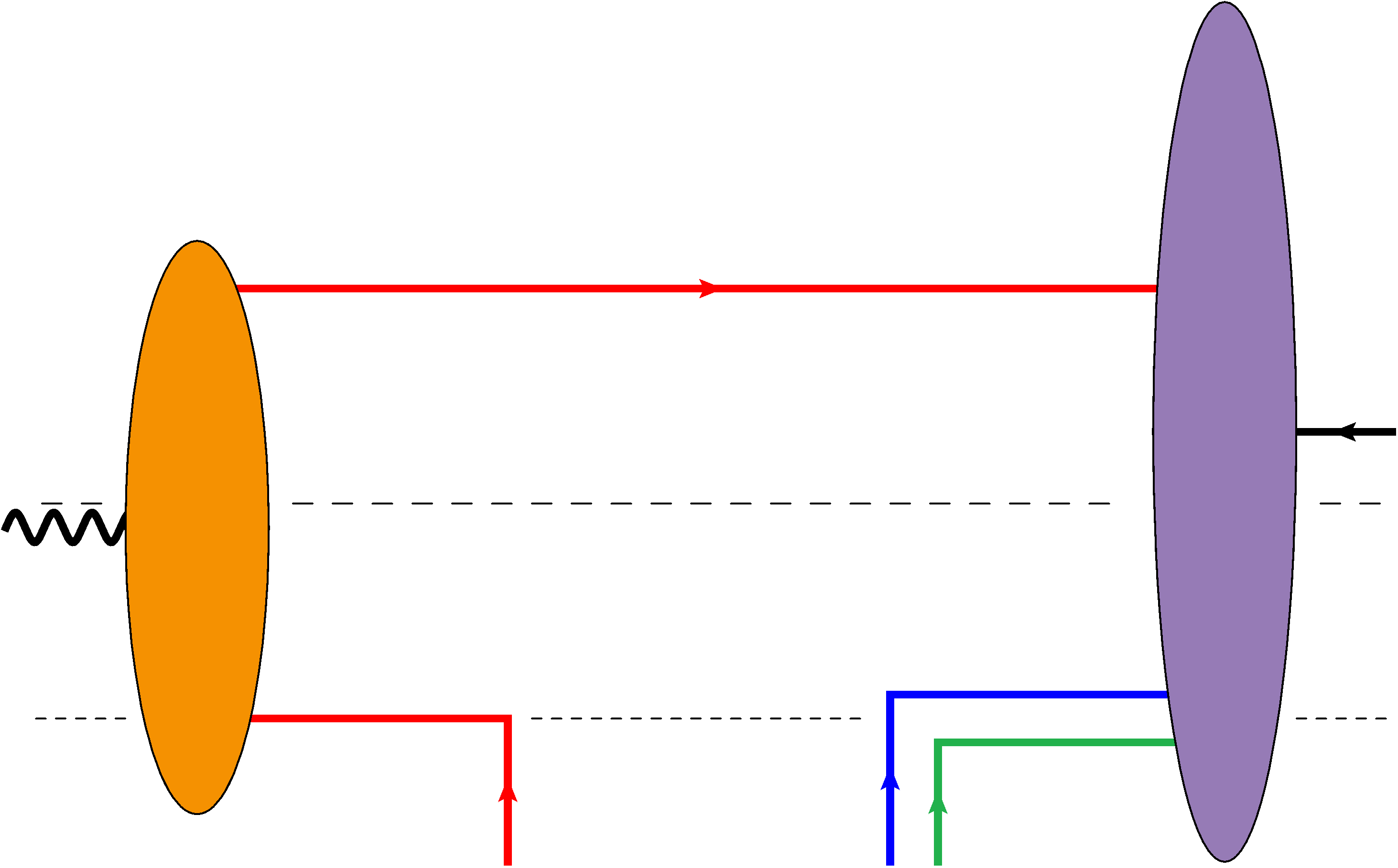}
\label{dglapone_fig}} \qquad \qquad \qquad \qquad \qquad \qquad
\subfloat[]{\includegraphics[width = 0.32\textwidth]{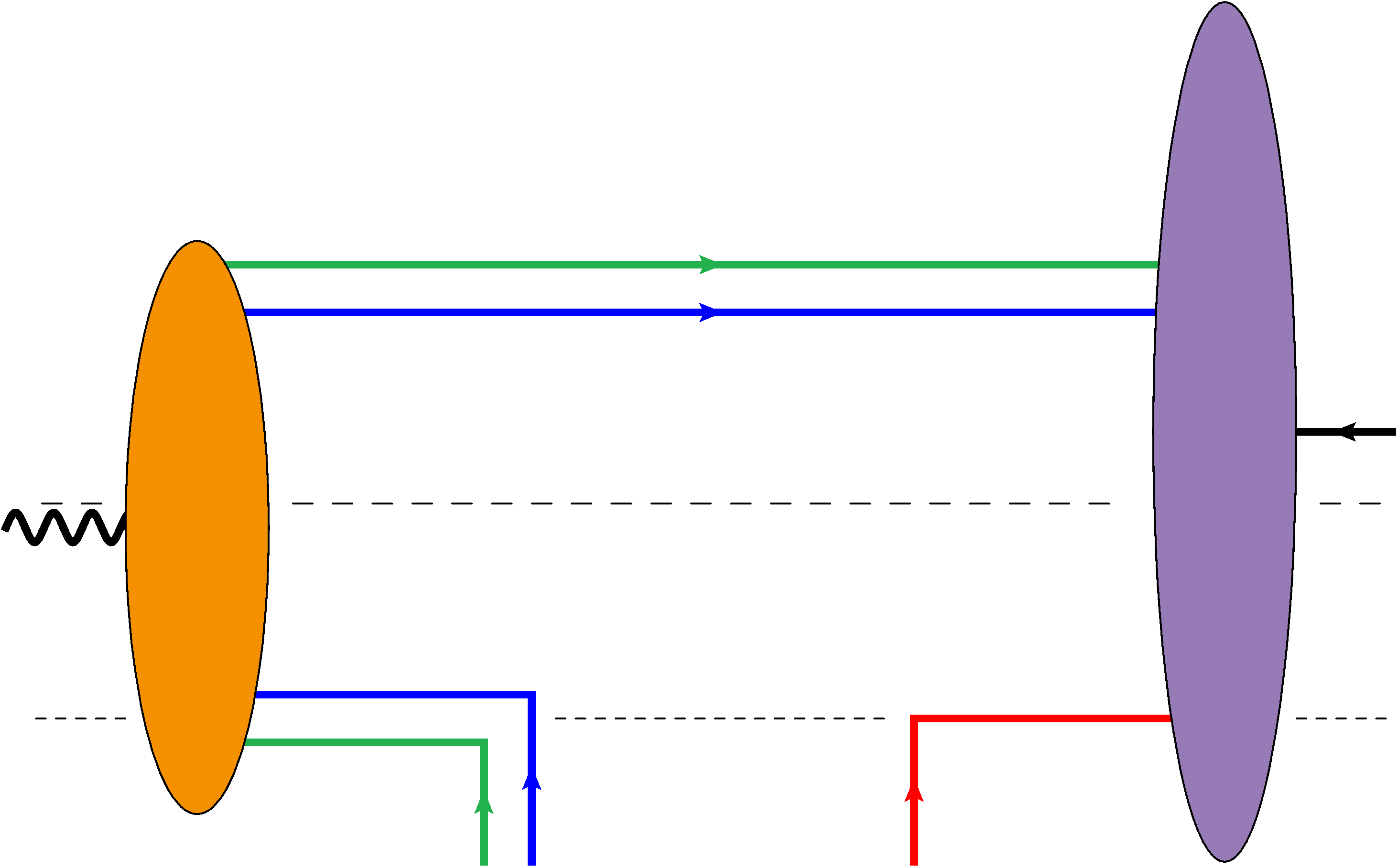}
\label{dglaptwo_fig}} \\
\subfloat[]{\includegraphics[width = 0.32\textwidth]{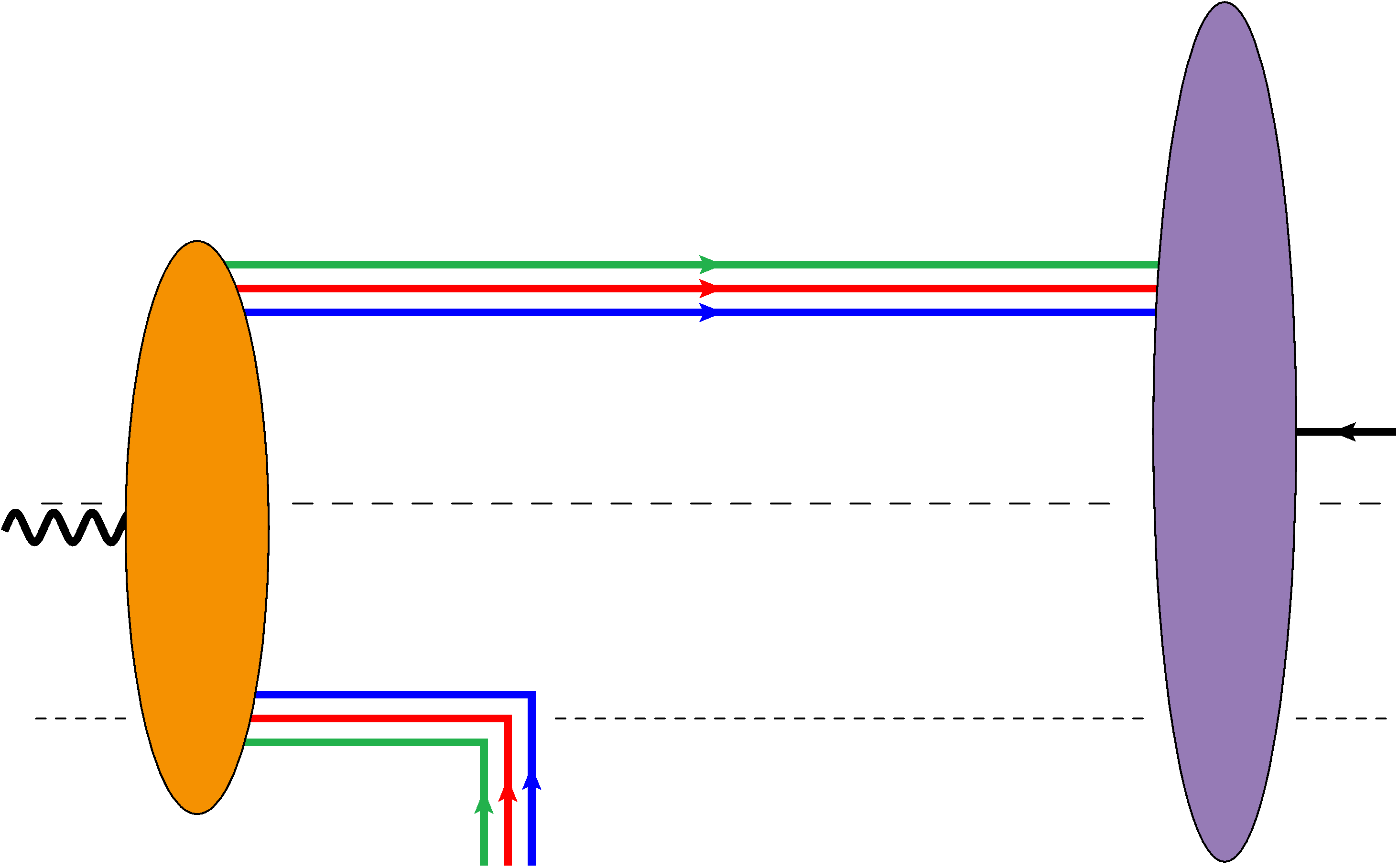}
\label{erbl_fig}}
\caption{Sketch of the leading contribution to photon-to-nucleon TDAs in the DGLAP I (a), DGLAP II (b), and ERBL (c) support regions.}
\label{tdas_support_rgns}
\end{figure}

In the context of collinear factorization, there are three fundamental ways in which the photon-to-nucleon transition can happen, schematically illustrated in Figs.~\ref{dglapone_fig}--\ref{erbl_fig}, where the photon is represented by the left (orange) oval shape and the nucleon by the right (violet) one. If the photon splits into a light (up or down) quark--antiquark pair, the antiquark is emitted and two absorbed quarks take its place (so that the result is a color singlet) to make up the nucleon (Fig.~\ref{dglapone_fig}). The photon could also split into two quark--antiquark pairs, with a quark taking the place of the antiquarks in the final nucleon (Fig.~\ref{dglaptwo_fig}), or into three pairs, the emitted antiquarks leaving behind the nucleon (Fig.~\ref{erbl_fig}). Additional quark--antiquark pairs (of any flavor) or gluons can be directly absorbed by the nucleon, and should give rise to higher-order corrections to the three fundamental contributions. The three classes of contributions are characterized by different values of the fractions of longitudinal mean momentum~\eqref{sudakov_mean} carried by the exchanged partons, with positive values corresponding to emitted antiquarks and negative values interpreted as absorbed quarks. The first and second support regions are called DGLAP I and DGLAP II, respectively, as they are governed by a generalization of the Dokshitzer--Gribov--Lipatov--Altarelli--Parisi evolution equations~\cite{Gribov:1972ri, Altarelli:1977zs, Dokshitzer:1977sg}. The third region is called ERBL, being controlled by a generalization of the Efremov--Radyushkin--Brodsky--Lepage evolution equations~\cite{Efremov:1978rn, Efremov:1979qk, Lepage:1979zb, Lepage:1979za, Lepage:1980fj}.

\subsection{Definition of the Photon-to-Nucleon Transition Distribution Amplitudes}

In order to define the photon-to-nucleon TDAs, we start by considering the matrix element of the trilocal operator that allows the transition between the initial photon and final nucleon, i.e., 
\begin{equation}
 \langle p_{N}', s_{N}' \lvert \Ohat_{ABC} \! \left( \lambda_{1}n, \lambda_{2}n, \lambda_{3}n \right) \rvert q, \lambda \rangle,
 \end{equation}
where $\lambda, s_{N}'$ are the LC helicities of the  photon and the proton, respectively, and the operator $\Ohat_{ABC}$ is defined as
 \begin{equation}
\Ohat_{ABC} \! \left( \lambda_{1}n, \lambda_{2}n, \lambda_{3}n \right) = \epsilon_{jkl} \overline{ \psi }_{u, jA} \! \left( \lambda_{1} n \right) \overline{ \psi }_{u, kB} \! \left( \lambda_{2} n \right) \overline{ \psi }_{d, lC} \! \left( \lambda_{3} n \right) \! .
\label{trilocal_operator_proton}
\end{equation}
In Eq.~\eqref{trilocal_operator_proton}, $ \overline{ \psi }_{u}, \overline{ \psi }_{d} $ are the Dirac adjoints of the field operators for up and down quarks, respectively, $A, B, C$ are Dirac indices, $j, k, l$ are summed (anti)quark-color indices, and $ \epsilon_{jkl} $ is the antisymmetric Levi-Civita symbol. For a final neutron, we just need to switch all up and down quarks, so we will only perform the calculations for the proton case. In general, we would have to insert Wilson lines between the local operators to ensure gauge invariance. However, choosing the LC gauge $A^{+} = 0$, these lines become trivial along the direction of $n$.

We Fourier transform the matrix element of the trilocal operator with the transformation
\begin{equation}
\mathcal{F} \! \left[ ... \right] = (p \cdot n)^{3} \! \int \! \left( \prod_{a = 1}^{3} \frac{d \lambda_{a} }{2 \pi} \right) \! \left[ ... \right] e^{i \sum_ {b = 1}^{3} x_{b} \lambda_{b} (p \cdot n)} \, .
\label{fourier_operator}
\end{equation}
In the LC decomposition of Eqs.~\eqref{sudakov_q}--\eqref{xi}, the result has the following structure:
\begin{align}
& \mathcal{F} \langle p_{N}', s_{N}' \lvert \Ohat_{ABC} \! \left( \lambda_{1}n, \lambda_{2}n, \lambda_{3}n \right) \rvert q, \lambda \rangle = \delta \! \left( x_{1} + x_{2} + x_{3} - 2 \xi \right) \frac{m_{N}}{4} \nonumber \\
& \times \sum_{ \mathcal{I} } \left( s_{ \mathcal{I} }^{N \gamma} \right)_{ \! ABC} \! \left( q, \lambda, p_{N}', s_{N}' \right) S_{ \mathcal{I} }^{N \gamma} \! \left( x_{1}, x_{2}, x_{3}, \xi, u \right) \! ,
\label{helicity_amp_tda_decompo}
\end{align}
where  the factor of $m_{N}/4$ is for convenience. The fractions of longitudinal momentum of the exchanged partons are in the following range:
\begin{equation}
\xi - 1 \leq x_{b} \leq \xi + 1, \quad b = 1, 2, 3.
\end{equation}
When $x_{b}$ is positive, it corresponds to  the fraction of released longitudinal momentum in the photon-to-proton transition carried by an emitted antiquark, while, when negative, its absolute value is the fraction carried by an absorbed quark. In agreement with this interpretation, the $x$s are constrained to add up to $2 \xi $. When one $x$ variable is positive and two are negative, we are in the DGLAP I support region (Fig.~\ref{dglapone_fig}), when two are positive and one is negative in the DGLAP II region (Fig.~\ref{dglaptwo_fig}), and when all three are positive in the ERBL region (Fig.~\ref{erbl_fig}). The sum in Eq.~(\ref{helicity_amp_tda_decompo}) is over a set of independent Dirac structures, and the coefficients are the photon-to-nucleon TDAs, which turn out to be dimensionless and real, and also depend on a collinear factorization scale. We will only consider the leading-twist TDAs, i.e., the contribution to the matrix element~\eqref{helicity_amp_tda_decompo} with the highest power of $P^{+}$. It comes from the LC good components of the spinors~\eqref{good_spinor}, and has a twist of three.

The Dirac structures are related to the ones for the nucleon-to-photon ($ \gamma N$) transition of DVCS, listed in Appendix B of 
Ref.~\cite{Pire:2022fbi}, by
\begin{equation}
\left( s^{N \gamma } \right)_{ \! ABC} = - \left( \gamma^{0 \top } \right)_{ \! AA'} \left( s^{ \gamma N \dagger } \right)_{ \! A'B'C'} \gamma^{0}_{B'B} \gamma^{0}_{C'C}.
\label{tcs_dirac_structs_from_dvcs}
\end{equation}
Note that our formula corrects the analogous Eq. (17) of Ref.~\cite{Pire:2022fbi}, which misses a minus sign. In exactly backward TCS, where $ \Delta_{T} = 0$, only four Dirac structures are nonzero. For convenience, we reproduce them below 
\begin{alignat}{1}
\left( v_{1 \mathcal{E} }^{ \gamma N} \right)_{ \! ABC}
& = \left( \phat \mathcal{C} \right)_{ \! AB} \left( \gamma^{5} \epsilonhat^{*} U_{+} \right)_{ \! C} \! ,
\label{v_onee_dvcs} \\
\left( a_{1 \mathcal{E} }^{ \gamma N} \right)_{ \! ABC}
& = \left( \phat \gamma^{5} \mathcal{C} \right)_{ \! AB} \left( \epsilonhat^{*} U_{+} \right)_{ \! C} \! ,
\label{a_onee_dvcs} \\
\left( t_{1 \mathcal{E} }^{ \gamma N} \right)_{ \! ABC}
& = \left( \sigma_{p \mu } \mathcal{C} \right)_{ \! AB} \left( \gamma^{5} \sigma^{\mu \epsilon^{*} } U_{+} \right)_{ \! C} \! ,
\label{t_onee_dvcs} \\
\left( t_{2 \mathcal{E} }^{ \gamma N} \right)_{ \! ABC}
& = \left( \sigma_{p \epsilon^{*} } \mathcal{C} \right)_{ \! AB} \left( \gamma^{5} U_{+} \right)_{ \! C} \! ,
\label{t_twoe_dvcs}
\end{alignat}
where $ \mathcal{C} $ is the charge conjugation matrix~\eqref{ks_chiral_charge_conjugation_matrix}, $ \epsilon^\mu $ is the photon polarization vector~\eqref{lcgauge_polarization_vectors}, $U_{+}$ is the good component of the nucleon spinor, and
\begin{alignat}{1}
\phat & = p_{ \nu } \gamma^{ \nu },
\label{hat_p} \\
\sigma^{ \nu \mu } & = \frac{1}{2} \left[ \gamma^{ \nu }, \gamma^{ \mu } \right] \! ,
\label{sigma_numu} \\
\sigma_{p \mu } & = p^{ \nu } \sigma_{ \nu \mu },
\label{sigma_pmu} \\
\sigma_{p \epsilon^{*} } & = p^{ \nu } \epsilon^{* \mu } \sigma_{ \nu \mu }.
\label{sigma_pepsilonstar}
\end{alignat}
If we define the helicity amplitudes 
\begin{align}
T^{s_{N}', \lambda }_{ABC} = \frac{m_{N}}{4} & \left( \left( v_{1 \mathcal{E}}^{N \gamma } \right)_{ \! ABC} V_{1 \mathcal{E}}^{N \gamma } + \left( a_{1 \mathcal{E} }^{N \gamma } \right)_{ \! ABC} A_{1 \mathcal{E}}^{N \gamma } \right. \nonumber \\
& \left. + \left( t_{1 \mathcal{E} }^{N \gamma } \right)_{ \! ABC} T_{1 \mathcal{E}}^{N \gamma } + \left( t_{2 \mathcal{E} }^{N \gamma } \right)_{ \! ABC} T_{2 \mathcal{E}}^{N \gamma } \right) \! ,
\label{helicity_amplitude_def}
\end{align}
we can express the TDAs as linear combinations of the helicity amplitudes , i.e.,
\begin{alignat}{1}
V_{1 \mathcal{E} }^{N \gamma}
& = \frac{ \sqrt{2} }{ \sqrt{1 - \xi } \left( P^{+} \right)^{ \! \frac{3}{2} } m_{N}} \left( T^{ \uparrow, +1}_{322} + T^{ \uparrow, +1}_{232} \right) \! ,
\label{bigv_one} \\
A_{1 \mathcal{E} }^{N \gamma}
& = \frac{ \sqrt{2} }{ \sqrt{1 - \xi } \left( P^{+} \right)^{ \! \frac{3}{2} } m_{N}} \left( T^{ \uparrow, +1}_{322} - T^{ \uparrow, +1}_{232} \right) \! ,
\label{biga_one} \\
T_{1 \mathcal{E} }^{N \gamma}
& = \frac{ \sqrt{2} }{ \sqrt{1 - \xi } \left( P^{+} \right)^{ \! \frac{3}{2} } m_{N}} \left( T^{ \uparrow, +1}_{223} - T^{ \uparrow, -1}_{333} \right) \! ,
\label{bigt_one} \\
T_{2 \mathcal{E} }^{N \gamma}
& = \frac{ \sqrt{2} }{ \sqrt{1 - \xi } \left( P^{+} \right)^{ \! \frac{3}{2} } m_{N}} \left( T^{ \uparrow, +1}_{223} + T^{ \uparrow, -1}_{333} \right) \! .
\label{bigt_two}
\end{alignat}
The choice of Dirac indices in the above equations fixes the components of the Dirac adjoints of the spinors in the expansions of the local fields in the trilocal operator~\eqref{trilocal_operator_proton}. Since we are focusing on the good components~\eqref{good_spinor}, an index value of 2 gives a nonzero contribution from a quark with spin $-1/2$ or antiquark with spin $+1/2$, while a quark with spin $+1/2$ or an antiquark with spin $-1/2$ contribute when the index is 3. Therefore, the helicity amplitudes in Eqs.~\eqref{bigv_one}--\eqref{bigt_two} correspond to transitions where the total LC helicity is conserved without any transfer of orbital angular momentum between the photon and the proton, which is compatible with $ \Delta_{T} = 0$. For example, considering $T^{ \uparrow, +1}_{322}$ and the photon splitting into a down-antidown pair, we have a photon and two up quarks of LC helicity $+1, +1/2, -1/2$, respectively, transitioning into a proton and an antidown both of helicity $+1/2$, for a conserved total helicity of $+1$. Note that the down quark could have helicity $+1/2$, corresponding to no orbital angular momentum between the partons in the photon nor in the proton, or it could have helicity $-1/2$, for a third component of orbital angular momentum of $+1$ both in the photon and in the proton. We can interpret the other cases in an analogous fashion.

\section{Modeling Photon-to-Nucleon Transition Distribution Amplitudes}
\label{section_model_tdas}

In the following, we focus on the study of the leading contribution to the TDAs in the DGLAP I region, corresponding to the probing of the $q \overline{q} $ and  $qqq$ components of the photon and proton states, respectively, as represented in Fig.~\ref{dglapone_fig}. We first introduce a model for the LFWFs of the photon and proton states, and then give the structure of the analytical results for the TDAs from the model calculation. We conclude with numerical predictions for a selected set of TDAs and their first Mellin moments.

\subsection{A Light-Front Dynamical Model for the Photon and the Nucleon}
\label{subsec_lfd_model}

In LFD, we can represent interacting states on a basis of Fock states, provided that we avoid particles with zero longitudinal momentum (see, e.g., Refs.~\cite{Brodsky:1997de, Keister:1991sb}). If $A$ labels a strongly interacting  particle with on-shell four-momentum $P$ and LC helicity $ \Lambda $, we can write
\begin{equation}
\rvert A; P, \Lambda \rangle = \sum_{N, \beta } \int \! \left[ \frac{dx}{ \sqrt{x} } \right]_{ \! N} \! \left[ d^{2}k_{ \perp } \right]_{ \! N} \Psi^{A, \Lambda}_{N, \beta} \left( \left\{ x_i, \vec{q}_{i \perp} \right\} \right) \rvert N; p_{1}, ..., p_{N}, \beta \rangle,
\label{fock_decompo}
\end{equation}
where $N = n_{g} + n_{q} + n_{ \overline{q} }$ labels the number of partons with discrete quantum numbers collectively denoted by $ \beta $. In Eq.~\eqref{fock_decompo}, the integration measures are defined as
\begin{alignat}{1}
\left[ \frac{dx}{ \sqrt{x} } \right]_{ \! N}
& = \prod_{i = 1}^{N} \frac{dx_{i}}{ \sqrt{x_{i}} } \delta \! \left( 1 - \sum_{j = 1}^{N} x_{j} \right) \! ,
\label{fock_decompo_xoversqrtxmeas} \\
\left[ d^{2}k_{ \perp } \right]_{ \! N}
& = \frac{1}{ \left( 2 \left( 2 \pi \right)^{3} \right)^{N - 1}} \prod_{i = 1}^{N} d^{2}k_{i \perp } \delta^{ \left( 2 \right) } \! \left( \sum_{j = 1}^{N} \vec{k}_{j \perp } \right) \! ,
\label{fock_decompo_relperpmeas}
\end{alignat}
where, for every parton with four-momentum $p_{i}$, we defined the fraction $x_i$ of longitudinal momentum with respect to $P^{+}$ and the transverse momentum $ \vec{k}_{i \perp } $ with respect to $ \vec{P}_{ \perp } $, i.e.,
\begin{alignat}{1}
p^{+}_{i}
& = x_{i}P^{+},
\label{longitmomentum_frac} \\
\vec{p}_{i \perp }
& = x_{i} \vec{P}_{ \perp } + \vec{k}_{ i\perp }.
\label{perpmomentum_frac}
\end{alignat}
The coefficients $ \Psi^{A, \Lambda }_{N, \beta } $ in Eq.~\eqref{fock_decompo} are the LFWFs that give the probability amplitude to find the $N$-parton Fock state in the hadron $A$. They
only depend on the relative variables  $\left( x_i, \vec{k}_{i \perp } \right) $ and are thus Lorentz invariant.

\begin{figure}[h]
\centering
\includegraphics[scale = 1.7]{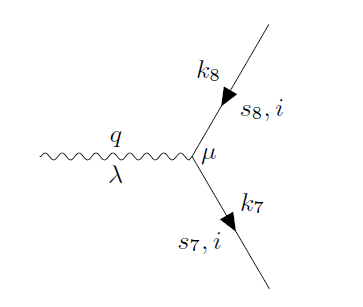}
\caption{Photon to quark--antiquark pair at tree level.}
\label{gammaqqbar_tree_diag}
\end{figure}

We model the initial photon as a quark--antiquark pair, with flavor either up ($u$) or down ($d$). The corresponding LFWF can be obtained  by the tree level diagram of Fig.~\ref{gammaqqbar_tree_diag} (see, e.g., Ref.~\cite{Kovchegov:2012mbw}). We have
\begin{align}
\Psi^{ \gamma, \lambda }_{f\bar f, 7 8} =
&-e_{f}e \sqrt{2} \left( \frac{m^{2} + k_{7 \perp }^{2}}{x_{7}} + \frac{m^{2} + k_{8 \perp }^{2}}{x_{8}} \right)^{-1} \nonumber \\
& \times \left( \frac{m}{x_{7}x_{8}} \delta_{ \lambda s_{7}} \delta_{s_{7}s_{8}} + \sqrt{2} \vec{\epsilon}_{ \lambda \perp } \cdot \left( \frac{ \vec{k}_{7 \perp }}{x_{7}} \delta_{ \lambda -s_{7}} \delta_{-s_{7}s_{8}} + \frac{ \vec{k}_{8 \perp } }{x_{8}} \delta_{ \lambda s_{7}} \delta_{s_{7}-s_{8}} \right) \right) \! ,
\label{gammaqqbar_wf_tree_rel_78}
\end{align}
where the subscripts $7, 8$  collectively denote the LC helicity and the momentum of quark and antiquark, respectively,  the Kronecker deltas only check the sign of the parton helicities, $e_{f}$ is the charge of quark flavor $f$ in units of the positron charge $e$, and $m$ is the light-quark mass. In Eq.~\eqref{gammaqqbar_wf_tree_rel_78}, $\vec{\epsilon}_{ \lambda \perp } $ is the transverse part of the  photon polarization vector, i.e.,
\begin{equation}
\epsilon_{ \lambda }^{ \mu } \! \left( q \right) = \left( 0, \frac{2 \vec{q}_{ \perp } \cdot \vec{ \epsilon }_{ \lambda \perp } }{q^{+}}, \vec{ \epsilon }_{ \lambda \perp } \right) \! , \qquad \vec{ \epsilon }_{ \lambda \perp } = - \frac{1}{ \sqrt{2} } \left( \lambda, i \right) \! ,
\label{lcgauge_polarization_vectors}
\end{equation}
with $ \lambda = +1, -1$ corresponding to counterclockwise and clockwise polarization, respectively.

For the proton, we use the constituent-quark LFWFs introduced in Ref.~\cite{Boffi:2002yy} and expanded in terms of eigenstates of orbital angular momentum  in  Ref.~\cite{Pasquini:2008ax}. The model was applied to calculate the orbital-angular-momentum content of transverse-momentum-dependent distributions~\cite{Pasquini:2008ax, Pasquini:2010af}, and to predict the spin asymmetries in semi-inclusive deep-inelastic scattering~\cite{Boffi:2009sh, Pasquini:2011tk}. The energy scale at which the nucleon is well represented by the constituent valence quarks was derived in Ref.~\cite{Pasquini:2011tk}, and is about $0.5 \, \text{GeV} $. 

For convenience, we report below the results for a proton with LC helicity $+1/2$, the only needed for Eqs.~\eqref{bigv_one}--\eqref{bigt_two}, numbering the three valence quarks as $4, 5, 6$. The proton state is given by
\begin{equation}
\vert p_{N}', \uparrow \rangle = \vert p_{N}', \uparrow \rangle_{L_{z} = 0} + \vert p_{N}', \uparrow \rangle_{L_{z} = +1} + \vert p_{N}', \uparrow \rangle_{L_{z} = -1} + \vert p_{N}', \uparrow \rangle_{L_{z} = +2},
\label{proton_valquarks_helicitybalance}
\end{equation}
where $L_{z}$ is the third component of the total orbital angular momentum of the three valence quarks. Every orbital-angular-momentum component corresponds to one of the possible combinations  of LC helicities of the quarks, so that the third component of total angular momentum $L_{z} + \sum_{i = 4, 5, 6} s_{i} = +1/2$. They are given by
\begin{alignat}{1}
\vert p_{N}', \uparrow \rangle_{0} ={}
& \int \! \left[ \frac{dx}{ \sqrt{x} } \right]_3 \! \left[ d^{2}k_{ \perp } \right]_3 \left( \psi^{ \left( 1 \right) } \! \left( 4, 5, 6 \right) + i \left( k_{4}^{1}k_{5}^{2} - k_{4}^{2}k_{5}^{1} \right) \psi^{ \left( 2 \right) } \! \left( 4, 5, 6 \right) \right) \nonumber \\
&\times  \frac{1}{ \sqrt{6} } \epsilon_{xyz} u_{\uparrow x}^{ \dagger }(4) \left( u_{\downarrow y}^{ \dagger }(5)d_{ \uparrow z}^{ \dagger }(6) - d_{\downarrow y}^{ \dagger }(5)u_{\uparrow z}^{ \dagger }(6) \right) \rvert 0 \rangle,
\label{proton_valquarks_lzzero} \\
\vert p_{N}', \uparrow \rangle_{+1} ={}
& \int \! \left[ \frac{dx}{ \sqrt{x} } \right]_3 \! \left[ d^{2}k_{ \perp } \right]_3 \left( k_{4 \perp }^{+} \psi^{ \left( 3 \right) } \! \left( 4, 5, 6 \right) + k_{5 \perp }^{+} \psi^{ \left( 4 \right) } \! \left( 4, 5, 6 \right) \right) \nonumber \\
& \times \frac{1}{ \sqrt{6} } \epsilon_{xyz} \left( u_{\uparrow x}^{ \dagger }(4)u_{\downarrow y}^{ \dagger }(5)d_{ \downarrow z}^{ \dagger }(6) - d_{\uparrow x}^{ \dagger }(4)u_{ \downarrow y}^{ \dagger }(5)u_{\downarrow z}^{ \dagger }(6) \right) \rvert 0 \rangle,
\label{proton_valquarks_lzone} \\
\vert p_{N}', \uparrow \rangle_{-1} ={}
& \int \! \left[ \frac{dx}{ \sqrt{x} } \right]_3 \! \left[ d^{2}k_{ \perp } \right]_3 \left( -1 \right)k_{5 \perp }^{-} \psi^{ \left( 5 \right) } \! \left( 4, 5, 6 \right) \nonumber \\
& \times \frac{1}{ \sqrt{6} } \epsilon_{xyz} u_{ \uparrow x}^{ \dagger } (4)\left( u_{\uparrow y}^{ \dagger }(5)d_{ \uparrow z}^{ \dagger }(6) - d_{\uparrow y}^{ \dagger }(5)u_{ \uparrow z}^{ \dagger }(6) \right) \rvert 0 \rangle,
\label{proton_valquarks_lzminone} \\
\vert p_{N}', \uparrow \rangle_{+2} ={}
& \int \! \left[ \frac{dx}{ \sqrt{x} } \right]_3 \! \left[ d^{2}k_{ \perp } \right]_3 \left( -1 \right) k_{4 \perp }^{+}k_{6 \perp }^{+} \psi^{ \left( 6 \right) } \! \left( 4, 5, 6 \right) \nonumber \\
& \times \frac{1}{ \sqrt{6} } \epsilon_{xyz} u_{ \downarrow x}^{ \dagger } (4)\left( u_{ \downarrow y}^{ \dagger }(5)d_{ \downarrow z}^{ \dagger }(6) - d_{ \downarrow y}^{ \dagger }(5)u_{\downarrow z}^{ \dagger (6)} \right) \rvert 0 \rangle,
\label{proton_valquarks_lztwo}
\end{alignat}
where $k^\pm_{i}=k^1_{i}\pm k^2_{i}.$ 
In Eqs.~\eqref{proton_valquarks_lzzero}--\eqref{proton_valquarks_lztwo}, $u^{ \dagger }_{s  x} \left( i \right) $ $ \left( d^{ \dagger }_{s x} \left( i \right) \right) $ is the creation operator for on-shell $u$ $ \left( d \right) $ quarks, with  LC helicity $s = \uparrow,\downarrow$ corresponding to $+1/2, -1/2$, respectively, $x$ is the color, and the argument $(i)$ stands for $ \left( x_{i}, k_{i \perp } \right) $. The functions $ \psi^{ \left( j \right) }$, with $j = 1, 2, ..., 6$, are given by (see Ref.~\cite{Pasquini:2008ax})
\begin{alignat}{1}
\psi^{ \left( 1 \right) } (4,5,6)={}
& \frac{1}{ \sqrt{3} } \psitilde \! \left( \left\{ x_{i}, \vec{k}_{i \perp } \right\} \right) \prod_{i} \frac{1}{ \sqrt{N_{i}} } \left( -a_{4}a_{5}a_{6} + \left( 2a_{4} + a_{6} \right) \vec{k}_{4 \perp } \cdot \vec{k}_{5 \perp } + 2a_{4}k_{5 \perp }^{2} \right) \! ,
\label{schlumpf_psi_one} \\
\psi^{ \left( 2 \right) } (4,5,6)={}
& \frac{1}{ \sqrt{3} } \psitilde \! \left( \left\{ x_{i}, \vec{k}_{i \perp } \right\} \right) \prod_{i}\frac{1}{ \sqrt{N_{i}} } \left( 2a_{4} + a_{6} \right) \! ,
\label{schlumpf_psi_two} \\
\psi^{ \left( 3 \right) } (4,5,6)={}
& \frac{1}{ \sqrt{3} } \psitilde \! \left( \left\{ x_{i}, \vec{k}_{i \perp } \right\} \right) \prod_{i} \frac{1}{ \sqrt{N_{i}} } \left( -a_{4}a_{5} - k_{5 \perp }^{2} \right) \! ,
\label{schlumpf_psi_three} \\
\psi^{ \left( 4 \right) } (4,5,6)={}
& \frac{1}{ \sqrt{3} } \psitilde \! \left( \left\{ x_{i}, \vec{k}_{i \perp } \right\} \right) \prod_{i} \frac{1}{ \sqrt{N_{i}} } \left( -a_{4}a_{5} - 2a_{4}a_{6} + k_{4 \perp }^{2} + 2 \vec{k}_{4 \perp } \cdot \vec{k}_{5 \perp } \right) \!,
\label{schlumpf_psi_four} \\
\psi^{ \left( 5 \right) } (4,5,6)={}
& \frac{1}{ \sqrt{3} } \psitilde \! \left( \left\{ x_{i}, \vec{k}_{i \perp } \right\} \right) \prod_{i} \frac{1}{ \sqrt{N_{i}} } a_{4}a_{6},
\label{schlumpf_psi_five} \\
\psi^{ \left( 6 \right) } (4,5,6)={}
& \frac{1}{ \sqrt{3} } \psitilde \! \left( \left\{ x_{i}, \vec{k}_{i \perp } \right\} \right) \prod_{i} \frac{1}{ \sqrt{N_{i}} } a_{5},
\label{schlumpf_psi_six}
\end{alignat}
where $N_{i} = a_{i}^{2} + k_{i \perp }^{2}$ and $ a_{i} = m + x_{i}M_{0}$, with $M_0$ the mass of the non-interacting three-quark system, i.e., $ M_{0} = \sqrt{ \sum_{i = 4}^{6} \left( m^{2} + k_{i \perp }^{2} \right) /x_{i} }.$
For the function $ \psitilde $, we  use
\begin{equation}
\psitilde \! \left( \left\{ x_{i}, \vec{k}_{i \perp } \right\} \right) = 2 \left( 2 \pi \right)^{3} \sqrt{ \frac{1}{M_{0}} \prod_{i = 4}^{6} \frac{ \omega_{i} }{x_{i}} } \frac{N'}{ \left( M_{0}^{2} + \betatilde^{2} \right)^{ \gammatilde } },
\label{schlumpf_psi} \\
\end{equation}
where $ N' $ is a normalization and $\omega_i = \left( m^{2} + k_{i \perp }^{2} \right) /x_{i}$ is the free-quark energy. The parameters $ \betatilde, \gammatilde $, as well as the light-quarks mass $m$, are taken from Ref.~\cite{Schlumpf:1992pp}, and are the result of a fit to the electromagnetic form factors. We have
\begin{equation}
m = 0.263 \, \text{GeV}, \qquad \betatilde = 0.607 \, \text{GeV}, \qquad \gammatilde = 3.5, \qquad \Ntilde = 0.047 \, \text{GeV}^{-4}.
\end{equation}

\subsection{Structure of the Transition Distribution Amplitudes}

We can now calculate the matrix elements of the trilocal operator in Eq.~\eqref{trilocal_operator_proton} by inserting the model for the photon and proton LFWFs described in Sec.~\ref{subsec_lfd_model}. Taking into account the quark-antiquark pair in the initial photon and the three valence quarks of the final nucleon, we end up evaluating matrix elements of the form
\begin{equation}
\langle q_{4x}q_{5y}q_{6z} \lvert \frac{1}{ \sqrt{6} } \epsilon_{xyz} \epsilon_{jkl} \overline{ \psi }_{u, jA} \! \left( \lambda_{1} n \right) \overline{ \psi }_{u, kB} \! \left( \lambda_{2} n \right) \overline{ \psi }_{d, lC} \! \left( \lambda_{3} n \right) \rvert q_{7i} \overline{q}_{8i} \rangle,
\label{qqbar_to_qqq}
\end{equation}
where $x, y, z$ are summed color indices and the index $i$ runs over color for the quark or the corresponding anticolor for the antiquark. Each number in the sets $4, 5, 6$ and $7, 8$ collectively denotes the other quantum numbers of a quark in the color-antisymmetric three-quark Fock component of the proton and in the quark--antiquark-pair Fock component of the photon, respectively. Focusing on the leading twist three and expanding the fields in terms of ladder operators, the only nonzero contributions come from annihilating an antiquark and creating two quarks between the initial and final states. We have two possibilities: the photon splits into a down--antidown pair ($d \overline{d} $) or into an up--antiup pair ($u \overline{u} $). In the first case, we annihilate the $ \overline{d} $ and create two $u$ and we are left with
\begin{align}
& \langle q_{4x}q_{5y}q_{6z} \lvert \frac{1}{ \sqrt{6} } \epsilon_{xyz} \Ohat_{ABC} \! \left( \lambda_{1}n, \lambda_{2}n, \lambda_{3}n \right) \rvert d_{7i} \overline{d}_{8i} \rangle \nonumber\\
&= - \sqrt{6}
\left(  \overline{u}_{u,s_{4}A} \! \left( k_{4} \right) \overline{u}_{u,s_{5}B} \! \left( k_{5} \right) e^{i \left( \lambda_{1} n \cdot k_{4} + \lambda_{2} n \cdot k_{5} \right)} \right.
 + \overline{u}_{u,s_{5}A} \! \left( k_{5} \right) \overline{u}_{u,s_{4}B} \! \left( k_{4} \right) e^{i \left( \lambda_{1} n \cdot k_{5} + \lambda_{2} n \cdot k_{4} \right)} \left. \vphantom{ \left( e^{i \left( \lambda_{1} n \cdot k_{5} - \lambda_{2} n \cdot  k_{8} \right)} \right) } \right) \nonumber \\
& \times \overline{v}_{d,s_{8}C} \! \left( k_{8} \right) e^{-i \lambda_{3} n \cdot k_{8}} \left( 2 \pi \right)^{3} 2k_{7}^{+} \delta_{s_{6}s_{7}} \delta \! \left( k_{6}^{+} - k_{7}^{+} \right) \delta^{\left( 2 \right)} \! \left( \vec{k}_{6 \perp} - \vec{k}_{7 \perp} \right) \! .
\label{ddbar_to_qqq} 
\end{align}
In the second case, we annihilate the $ \overline{u} $ and create $u, d$, so that
\begin{align}
& \langle q_{4x}q_{5y}q_{6z} \lvert \frac{1}{ \sqrt{6} } \epsilon_{xyz} \Ohat_{ABC} \! \left( \lambda_{1}n, \lambda_{2}n, \lambda_{3}n \right) \rvert u_{7i} \overline{u}_{8i} \rangle \nonumber\\
&= - \sqrt{6}  
\left( \overline{v}_{u,s_{8}A} \! \left( k_{8} \right) \overline{u}_{u,s_{5}B} \! \left( k_{5} \right) e^{-i \left( \lambda_{1} n \cdot k_{8} - \lambda_{2} n \cdot k_{5} \right)}
+ \overline{u}_{u,s_{5}A} \! \left( k_{5} \right) \overline{v}_{u,s_{8}B} \! \left( k_{8} \right) e^{+i \left( \lambda_{1} n \cdot k_{5} - \lambda_{2} n \cdot k_{8} \right)}  \right) \nonumber \\
& \times \overline{u}_{d, s_{6}C} \! \left( k_{6} \right) e^{i \lambda_{3} n \cdot k_{6}} \left( 2 \pi \right)^{3} 2k_{7}^{+} \delta_{s_{4}s_{7}} \delta \! \left( k_{4}^{+} - k_{7}^{+} \right) \delta^{\left( 2 \right)} \! \left( \overrightarrow{k}_{4 \perp} - \overrightarrow{k}_{7 \perp} \right) \nonumber \\
&- \sqrt{6}  \left(\overline{v}_{u,s_{8}A} \! \left( k_{8} \right) \overline{u}_{u,s_{4}B} \! \left( k_{4} \right) e^{-i \left( \lambda_{1} n \cdot k_{8} - \lambda_{2} n \cdot k_{4} \right)}
+   
\overline{u}_{u,s_{4}A} \! \left( k_{4} \right) \overline{v}_{u,s_{8}B} \! \left( k_{8} \right) e^{+i \left( \lambda_{1} n \cdot k_{4} - \lambda_{2} n \cdot k_{8} \right)}  \right) \nonumber \\
& \times \overline{u}_{d,s_{6}C} \! \left( k_{6} \right) e^{i \lambda_{3} n \cdot k_{6}} \left( 2 \pi \right)^{3} 2k_{7}^{+} \delta_{s_{5}s_{7}} \delta \! \left( k_{5}^{+} - k_{7}^{+} \right) \delta^{\left( 2 \right)} \! \left( \overrightarrow{k}_{5 \perp} - \overrightarrow{k}_{7 \perp} \right) \left. \vphantom{ \left( e^{- \frac{i}{2} \left( \lambda_{2} \tilde{y}_{8} - \lambda_{1} \tilde{y}_{5} \right)} \right) \delta^{\left( 2 \right)} \left( \overrightarrow{k}_{4 \perp} - \overrightarrow{k}_{7 \perp} \right) } \right) \! .
\label{uubar_to_qqq}
\end{align}

For the contribution from the proton LFWFs, we introduce the following notation for the a proton with LC helicity $+1/2$ $ \left( \uparrow \right) $ made up of three quarks:
\begin{alignat}{2}
& \uparrow \rightarrow \uparrow \downarrow \uparrow
& \text{ + }
& 0 = \psi^{ \left( 1 \right) } \! \left( 4, 5, 6 \right) + i \left( k_{4}^{1}k_{5}^{2} - k_{4}^{2}k_{5}^{1} \right) \psi^{ \left( 2 \right) } \! \left( 4, 5, 6 \right) \! ,
\label{p_into_pmp_zero} \\
& \uparrow \rightarrow \uparrow \downarrow \downarrow
& \text{ + }
& 1 = k_{4 \perp }^{+} \psi^{ \left( 3 \right) } \! \left( 4, 5, 6 \right) + k_{5 \perp }^{+} \psi^{ \left( 4 \right) } \! \left( 4, 5, 6 \right) \! ,
\label{p_into_pmm_plusone} \\
& \uparrow \rightarrow \uparrow \uparrow \uparrow
& \text{ $-$ }
& 1 = \left( -1 \right)k_{5 \perp }^{-} \psi^{ \left( 5 \right) } \! \left( 4, 5, 6 \right) \! ,
\label{p_into_ppp_minusone} \\
& \uparrow \rightarrow \downarrow \downarrow \downarrow
& \text{ + }
& 2 = \left( -1 \right) k_{4 \perp }^{+}k_{6 \perp }^{+} \psi^{ \left( 6 \right) } \! \left( 4, 5, 6 \right) \! ,
\label{p_into_mmm_plustwo}
\end{alignat}
where we have explicitly specified the total quark orbital angular momentum $L_{z} = 0,\pm 1, +2$, which combines with the quark LC helicities to give the proton LC helicity.

\begin{figure}[h]
\centering
	\resizebox*{!}{0.67\textheight}
	{
    		\begin{forest}
    		forked edges,
        	for tree =
        	{
        		anchor = center,
        		grow' = east,
            	draw,
            	rounded corners,
        	}
        	[{\hyperref[gammaqqbar_wf_tree_rel_78]{$ \Psi^{ \gamma, \lambda }_{d \overline{d}, 78} $}},
        		[$ \rvert d_{4}u_{5}u_{6} \rangle $, fill = Aquamarine,
            		[$ \overline{u}_{u, 5A} \overline{u}_{u, 6B} \overline{v}_{d, 8C} $,
                		[{\hyperref[p_into_pmm_plusone]{$ \uparrow \rightarrow \uparrow \downarrow \downarrow $ + 1}}, draw = Aquamarine]
                	]
                	[$ \overline{u}_{u, 6A} \overline{u}_{u, 5B} \overline{v}_{d, 8C} $,
                    	[{\hyperref[p_into_pmm_plusone]{$ \uparrow \rightarrow \uparrow \downarrow \downarrow $ + 1}}, draw = Aquamarine]
                	]
            	]
           	[$ \rvert u_{4}d_{5}u_{6} \rangle $, fill = DarkOrchid,
                	[$ \overline{u}_{u, 4A} \overline{u}_{u, 6B} \overline{v}_{d, 8C} $,
                    	[{\hyperref[p_into_pmp_zero]{$ \uparrow \rightarrow \uparrow \downarrow \uparrow $ + 0}}, draw = DarkOrchid]
                    	[{\hyperref[p_into_ppp_minusone]{$ \uparrow \rightarrow \uparrow \uparrow \uparrow $ $-$ 1}}, draw = DarkOrchid]
                    	[{\hyperref[p_into_mmm_plustwo]{$ \uparrow \rightarrow \downarrow \downarrow \downarrow $ + 2}}, draw = DarkOrchid]
                	]
                	[$ \overline{u}_{u, 6A} \overline{u}_{u, 4B} \overline{v}_{d, 8C} $,
                    	[{\hyperref[p_into_pmp_zero]{$ \uparrow \rightarrow \uparrow \downarrow \uparrow $ + 0}}, draw = DarkOrchid]
                    	[{\hyperref[p_into_ppp_minusone]{$ \uparrow \rightarrow \uparrow \uparrow \uparrow $ $-$ 1}}, draw = DarkOrchid]
                    	[{\hyperref[p_into_mmm_plustwo]{$ \uparrow \rightarrow \downarrow \downarrow \downarrow $ + 2}}, draw = DarkOrchid]
                	]
            	]
            	[$ \rvert u_{4}u_{5}d_{6} \rangle $, fill = BurntOrange,
            		[$ \overline{u}_{u, 4A} \overline{u}_{u, 5B} \overline{v}_{d, 8C} $,
                    	[{\hyperref[p_into_pmp_zero]{$ \uparrow \rightarrow \uparrow \downarrow \uparrow $ + 0}} , draw = BurntOrange]
                    	[{\hyperref[p_into_pmm_plusone]{$ \uparrow \rightarrow \uparrow \downarrow \downarrow $ + 1}}, draw = BurntOrange]
                    	[{\hyperref[p_into_ppp_minusone]{$ \uparrow \rightarrow \uparrow \uparrow \uparrow $ $-$ 1}}, draw = BurntOrange]
                    	[{\hyperref[p_into_mmm_plustwo]{$ \uparrow \rightarrow \downarrow \downarrow \downarrow $ + 2}}, draw = BurntOrange]
                	]
                	[$ \overline{u}_{u, 5A} \overline{u}_{u, 4B} \overline{v}_{d, 8C} $,
                    	[{\hyperref[p_into_pmp_zero]{$ \uparrow \rightarrow \uparrow \downarrow \uparrow $ + 0}}, draw = BurntOrange]
                    	[{\hyperref[p_into_pmm_plusone]{$ \uparrow \rightarrow \uparrow \downarrow \downarrow $ + 1}}, draw = BurntOrange]
                    	[{\hyperref[p_into_ppp_minusone]{$ \uparrow \rightarrow \uparrow \uparrow \uparrow $ $-$ 1}}, draw = BurntOrange]
                    	[{\hyperref[p_into_mmm_plustwo]{$ \uparrow \rightarrow \downarrow \downarrow \downarrow $ + 2}}, draw = BurntOrange]
                	]
            	]
        ]
                \end{forest}
	}

\caption{(Color online.) Structure of the helicity amplitude $T^{ \uparrow, \lambda}_{ABC}$ with $ \gamma \rightarrow d \overline{d}$. }
\label{hamplitude_from_ddbar}
\end{figure}
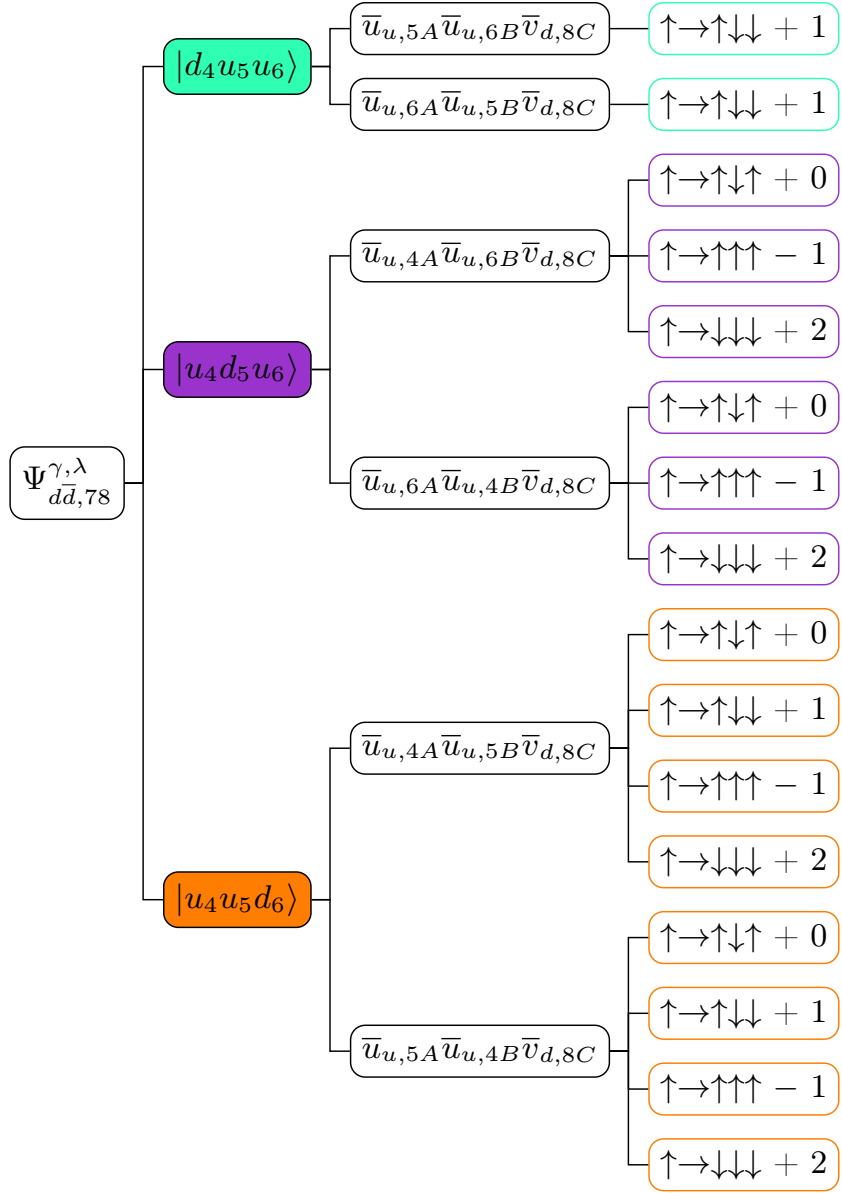

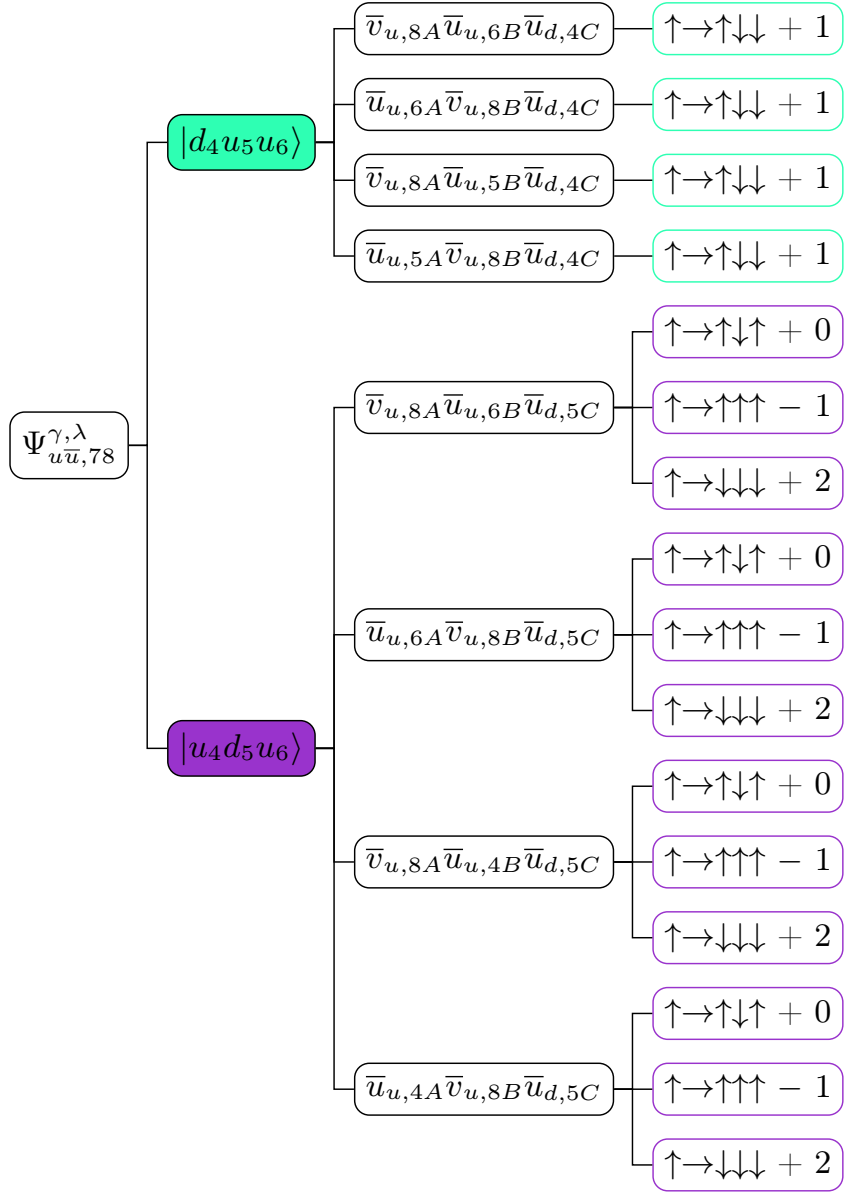
\begin{figure}[h]
\centering
	\resizebox*{!}{0.67\textheight}
	{
    		\begin{forest}
    		forked edges,
        	for tree =
        	{
        		anchor = center,
        		grow' = east,
            	draw,
            	rounded corners,
        	}
        	[{\hyperref[gammaqqbar_wf_tree_rel_78]{$ \Psi^{ \gamma, \lambda }_{u \overline{u}, 78} $}},
        		[$ \rvert d_{4}u_{5}u_{6} \rangle $, fill = Aquamarine,
            		[$ \overline{v}_{u, 8A} \overline{u}_{u, 6B} \overline{u}_{d, 4C} $,
                		[{\hyperref[p_into_pmm_plusone]{$ \uparrow \rightarrow \uparrow \downarrow \downarrow $ + 1}}, draw = Aquamarine]
                	]
                	[$ \overline{u}_{u, 6A} \overline{v}_{u, 8B} \overline{u}_{d, 4C} $,
                 	[{\hyperref[p_into_pmm_plusone]{$ \uparrow \rightarrow \uparrow \downarrow \downarrow $ + 1}}, draw = Aquamarine]
                	]
                	[$ \overline{v}_{u, 8A} \overline{u}_{u, 5B} \overline{u}_{d, 4C} $,
                 	[{\hyperref[p_into_pmm_plusone]{$ \uparrow \rightarrow \uparrow \downarrow \downarrow $ + 1}}, draw = Aquamarine]
                	]
                	[$ \overline{u}_{u, 5A} \overline{v}_{u, 8B} \overline{u}_{d, 4C} $,
                		[{\hyperref[p_into_pmm_plusone]{$ \uparrow \rightarrow \uparrow \downarrow \downarrow $ + 1}}, draw = Aquamarine]
                	]
            	]
           	[$ \rvert u_{4}d_{5}u_{6} \rangle $, fill = DarkOrchid,
            		[$ \overline{v}_{u, 8A} \overline{u}_{u, 6B} \overline{u}_{d, 5C} $,
                		[{\hyperref[p_into_pmp_zero]{$ \uparrow \rightarrow \uparrow \downarrow \uparrow $ + 0}}, draw = DarkOrchid]
                    	[{\hyperref[p_into_ppp_minusone]{$ \uparrow \rightarrow \uparrow \uparrow \uparrow $ $-$ 1}}, draw = DarkOrchid]
                    	[{\hyperref[p_into_mmm_plustwo]{$ \uparrow \rightarrow \downarrow \downarrow \downarrow $ + 2}}, draw = DarkOrchid]
               	]
                	[$ \overline{u}_{u, 6A} \overline{v}_{u, 8B} \overline{u}_{d, 5C} $,
                		[{\hyperref[p_into_pmp_zero]{$ \uparrow \rightarrow \uparrow \downarrow \uparrow $ + 0}}, draw = DarkOrchid]
                    	[{\hyperref[p_into_ppp_minusone]{$ \uparrow \rightarrow \uparrow \uparrow \uparrow $ $-$ 1}}, draw = DarkOrchid]
                    	[{\hyperref[p_into_mmm_plustwo]{$ \uparrow \rightarrow \downarrow \downarrow \downarrow $ + 2}}, draw = DarkOrchid]
                	]
                	[$ \overline{v}_{u, 8A} \overline{u}_{u, 4B} \overline{u}_{d, 5C} $,
                    	[{\hyperref[p_into_pmp_zero]{$ \uparrow \rightarrow \uparrow \downarrow \uparrow $ + 0}} , draw = DarkOrchid]
                    	[{\hyperref[p_into_ppp_minusone]{$ \uparrow \rightarrow \uparrow \uparrow \uparrow $ $-$ 1}}, draw = DarkOrchid]
                    	[{\hyperref[p_into_mmm_plustwo]{$ \uparrow \rightarrow \downarrow \downarrow \downarrow $ + 2}}, draw = DarkOrchid]
                	]
                	[$ \overline{u}_{u, 4A} \overline{v}_{u, 8B} \overline{u}_{d, 5C} $,
                    	[{\hyperref[p_into_pmp_zero]{$ \uparrow \rightarrow \uparrow \downarrow \uparrow $ + 0}}, draw = DarkOrchid]
                    	[{\hyperref[p_into_ppp_minusone]{$ \uparrow \rightarrow \uparrow \uparrow \uparrow $ $-$ 1}}, draw = DarkOrchid]
                    	[{\hyperref[p_into_mmm_plustwo]{$ \uparrow \rightarrow \downarrow \downarrow \downarrow $ + 2}}, draw = DarkOrchid]
                	]
            	]
        ]
                \end{forest}
	}
\caption{(Color online.) Structure of the helicity amplitude $T^{ \uparrow, \lambda}_{ABC}$ with $ \gamma \rightarrow u \overline{u}$, part I. }
\label{hamplitude_from_uubar_partone}
\end{figure}

\begin{figure}[h]
\centering
	\resizebox*{!}{0.67\textheight}
	{
    		\begin{forest}
    		forked edges,
        	for tree =
        	{
        		anchor = center,
        		grow' = east,
            	draw,
            	rounded corners,
        	}
        	[{\hyperref[gammaqqbar_wf_tree_rel_78]{$ \Psi^{ \gamma, \lambda }_{u \overline{u}, 78} $}},
        		[$ \rvert u_{4}u_{5}d_{6} \rangle $, fill = BurntOrange,
            		[$ \overline{v}_{u, 8A} \overline{u}_{u, 5B} \overline{u}_{d, 6C} $,
                		[{\hyperref[p_into_pmp_zero]{$ \uparrow \rightarrow \uparrow \downarrow \uparrow $ + 0}} , draw = BurntOrange]
                    	[{\hyperref[p_into_pmm_plusone]{$ \uparrow \rightarrow \uparrow \downarrow \downarrow $ + 1}}, draw = BurntOrange]
                    	[{\hyperref[p_into_ppp_minusone]{$ \uparrow \rightarrow \uparrow \uparrow \uparrow $ $-$ 1}}, draw = BurntOrange]
                    	[{\hyperref[p_into_mmm_plustwo]{$ \uparrow \rightarrow \downarrow \downarrow \downarrow $ + 2}}, draw = BurntOrange]
               	]
                	[$ \overline{u}_{u, 5A} \overline{v}_{u, 8B} \overline{u}_{d, 6C} $,
                    	[{\hyperref[p_into_pmp_zero]{$ \uparrow \rightarrow \uparrow \downarrow \uparrow $ + 0}}, draw = BurntOrange]
                    	[{\hyperref[p_into_pmm_plusone]{$ \uparrow \rightarrow \uparrow \downarrow \downarrow $ + 1}}, draw = BurntOrange]
                    	[{\hyperref[p_into_ppp_minusone]{$ \uparrow \rightarrow \uparrow \uparrow \uparrow $ $-$ 1}}, draw = BurntOrange]
                    	[{\hyperref[p_into_mmm_plustwo]{$ \uparrow \rightarrow \downarrow \downarrow \downarrow $ + 2}}, draw = BurntOrange]
                	]
            		[$ \overline{v}_{u, 8A} \overline{u}_{u, 4B} \overline{u}_{d, 6C} $,
                		[{\hyperref[p_into_pmp_zero]{$ \uparrow \rightarrow \uparrow \downarrow \uparrow $ + 0}} , draw = BurntOrange]
                    	[{\hyperref[p_into_pmm_plusone]{$ \uparrow \rightarrow \uparrow \downarrow \downarrow $ + 1}}, draw = BurntOrange]
                    	[{\hyperref[p_into_ppp_minusone]{$ \uparrow \rightarrow \uparrow \uparrow \uparrow $ $-$ 1}}, draw = BurntOrange]
                    	[{\hyperref[p_into_mmm_plustwo]{$ \uparrow \rightarrow \downarrow \downarrow \downarrow $ + 2}}, draw = BurntOrange]
               	]
                	[$ \overline{u}_{u, 4A} \overline{v}_{u, 8B} \overline{u}_{d, 6C} $,
                    	[{\hyperref[p_into_pmp_zero]{$ \uparrow \rightarrow \uparrow \downarrow \uparrow $ + 0}}, draw = BurntOrange]
                    	[{\hyperref[p_into_pmm_plusone]{$ \uparrow \rightarrow \uparrow \downarrow \downarrow $ + 1}}, draw = BurntOrange]
                    	[{\hyperref[p_into_ppp_minusone]{$ \uparrow \rightarrow \uparrow \uparrow \uparrow $ $-$ 1}}, draw = BurntOrange]
                    	[{\hyperref[p_into_mmm_plustwo]{$ \uparrow \rightarrow \downarrow \downarrow \downarrow $ + 2}}, draw = BurntOrange]
                	]
           	]
        ]
                \end{forest}
	}
\caption{(Color online.) Structure of the helicity amplitude $T^{ \uparrow, \lambda}_{ABC}$ with $ \gamma \rightarrow u \overline{u}$, part II. }
\label{hamplitude_from_uubar_partwo}
\end{figure}
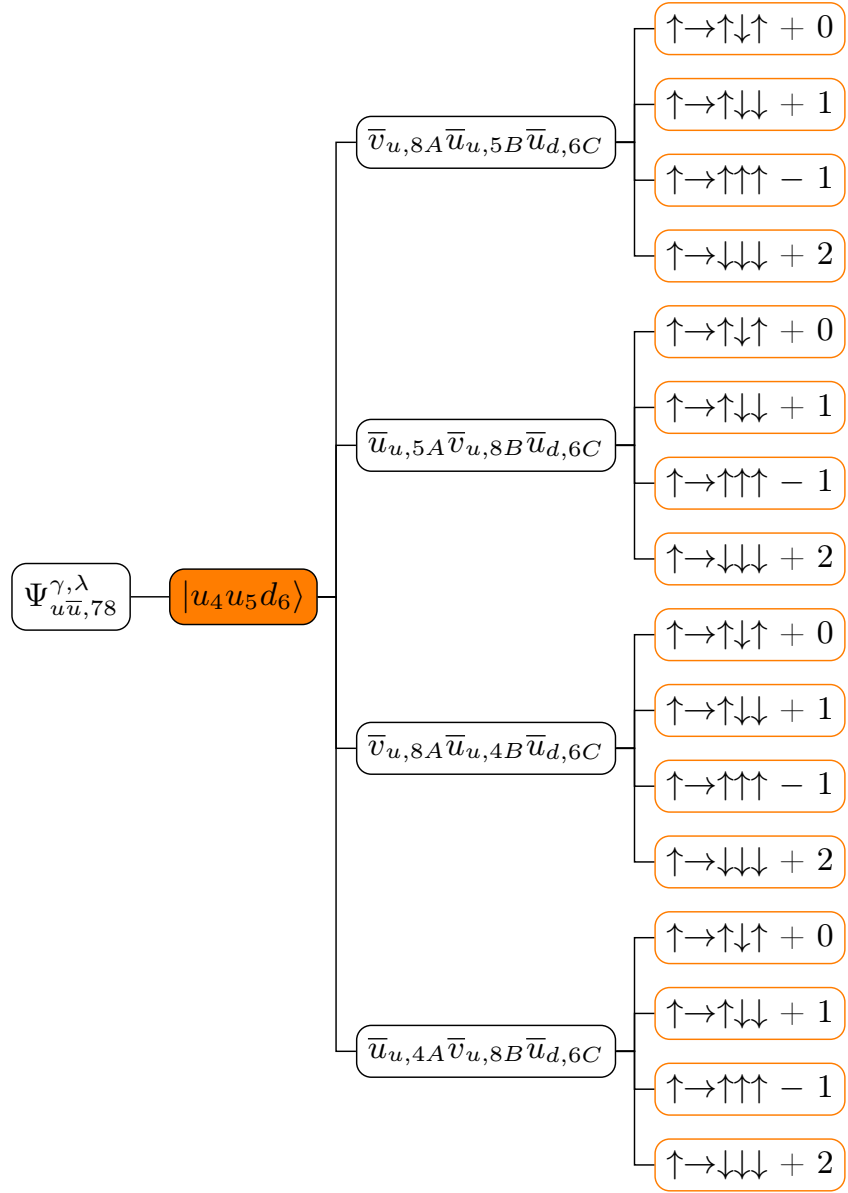

To obtain the final expressions for the photon-to-nucleon TDAs, we need to combine Eqs.~\eqref{ddbar_to_qqq} or \eqref{uubar_to_qqq} and Eq.~\eqref{gammaqqbar_wf_tree_rel_78} with Eqs.~\eqref{p_into_pmp_zero}--\eqref{p_into_mmm_plustwo}, and to sum and integrate over the quantum numbers of intermediate partons. Depending on the helicity amplitude~\eqref{helicity_amplitude_def} that we want to calculate, the result is a sum of many terms that we find impractical to write out in full. Instead, we schematically represent all the terms for a generic helicity amplitude in Figs.~\ref{hamplitude_from_ddbar}--\ref{hamplitude_from_uubar_partwo}, and we list below the steps to reconstruct the photon-to-proton TDAs of backward TCS.
\begin{itemize}
\item[1.] A term of an helicity amplitude is given by a path in Figs.~\ref{hamplitude_from_ddbar}--\ref{hamplitude_from_uubar_partwo} by multiplying the nodes in the first and third columns with the complex conjugate of the fourth-column one. The triplet of arrows in a fourth-column node represents the LC helicities of the three quarks in the final proton, in the order of the corresponding second-column node. We add a third component of orbital angular momentum to the LC helicities of the quarks, so that the total is the $+1/2$ LC helicity of the proton. The third-column nodes with flavor order $duu$ and $udu$ have been obtained from Eqs.~\eqref{ddbar_to_qqq}, \eqref{uubar_to_qqq} by the appropriate exchange of indices. Note that we are only considering the Dirac adjoint of the good components of the LC spinors~\eqref{good_spinor}. 
\item[2.] Fix the Dirac indices $A, B, C$. In the chiral representation~\eqref{ks_chiral_gammatrices}, the Dirac adjoints of the good components of the LC spinors~\eqref{good_spinor} are nonzero only for Dirac indices equal to $2, 3$, corresponding to LC helicity $-1/2, +1/2$ for particles, respectively, and helicity $+1/2, -1/2$ for antiparticles, respectively. 
\item[3.] We multiply by a factor of $ \prod_{i = 1}^{3} \exp \left[ \pm i \left( \left( \lambda_{i}/2 \right) x \right) \right] $, with $ \lambda_{1}, \lambda_{2}, \lambda_{3} $ associated with a spinor in the third-column node, in the order of the node, and with the plus sign for particles and the minus sign for antiparticles. We also multiply by a factor of the form
\begin{equation}
\left( 2 \pi \right)^{3} 2 k_{7}^{+} \delta_{s_{j}s_{7}} \delta \! \left( k_{j}^{+} - k_{7}^{+} \right) \delta^{ \left( 2 \right) } \! \left( \vec{k}_{j \perp } - \vec{k}_{7 \perp } \right) \! ,
\label{deltas_and_co_7protonquark}
\end{equation}
with $j$ equal to the value between $4, 5, 6$ that does not appear in the third-column node that we are considering.
\item[4.] We introduce a sum over the LC helicities of the quark--antiquark pair in the initial photon, and integrate over the three quark momenta in the proton and the quark and antiquark momenta in the photon with the measures~\eqref{fock_decompo_xoversqrtxmeas}--\eqref{fock_decompo_relperpmeas}. We then Fourier transform with the transformation~\eqref{fourier_operator}, which results in three Dirac delta distributions for the $x$ variables (see the following item).
\item[5.] We use all the Dirac delta distributions and Kronecker deltas to fix all but two longitudinal momenta, that we always choose to be index 4 and 5. For every term of an helicity amplitude, we have
\begin{equation}
\vec{k}_{6 \perp } = - \vec{k}_{4 \perp } - \vec{k}_{5 \perp }, \qquad x_{7} = 1 - x_{8}, \qquad \vec{k}_{8 \perp } = - \vec{k}_{7 \perp }.
\label{fixed_variables_for_all_helicity_amplitudes}
\end{equation}
The other variables for every surviving path of the helicity amplitudes that we are interested in are given by the rows of Tabs.~\ref{tthreetwotwo_tab}--\ref{tthreethreethree_tab_partwo}. Note that, by Eq.~\eqref{gammaqqbar_wf_tree_rel_78}, the photon cannot split into two partons whose helicities are both of the opposite sign of the original helicity, which would correspond to the $L_{z} = \pm 2$ components for the photon state. Therefore, the configurations with $s_{7} = s_{8} = -1/2$ do not appear in Tabs.~\ref{tthreetwotwo_tab}--\ref{ttwotwothree_tab}, and the entries with $s_{7} = s_{8} = +1/2$ do not appear in Tabs.~\ref{tthreethreethree_tab_partone}, \ref{tthreethreethree_tab_partwo}. Also note that the $L_{z} = +1$ component only contributes to $T^{ \uparrow, +1}_{322},  T^{ \uparrow, +1}_{232}, T^{ \uparrow, +1}_{223}$, while the $L_{z} = -1$ component only contributes to $T^{ \uparrow, -1}_{333}$.
\item[6.] For every term, we have a factor of 
\begin{equation}
\delta \! \left( x_{1} + x_{2} + x_{3} - 2 \xi \right) \! ,
\label{delta_of_extracted_pmomemntum}
\end{equation}
which is consistent with the fact that we have to extract a total of $2 \xi $ in going from the initial photon with momentum~\eqref{sudakov_q} to the final proton with momentum~\eqref{sudakov_pnprime}. This Dirac delta distribution is not included in the definition of the helicity amplitudes.
\item[7.] We add all the terms for an helicity amplitude, we add or subtract the helicity amplitudes according to Eqs.~\eqref{bigv_one}--\eqref{bigt_two}, and multiply by the remaining factor of
\begin{equation}
\frac{- \sqrt{6} \sqrt{2} }{m_{N} \left( 1 - \xi \right) \left( P^{+} \right)^{ \frac{3}{2} } },
\label{tdas_final_factor}
\end{equation}
with $P^{+}$ as in Eq.~\eqref{sudakov_mean} (Note that the factors of $P^{+}$ cancel overall).
\end{itemize}
\newpage
\vspace*{\fill}
\vspace*{\fill}
\newpage
\vspace*{\fill}
\vspace*{\fill}
\newpage
\vspace*{\fill}
\begin{table}[H]
\centering
\begin{NiceTabular}{ccccccc}[hvlines, name = t322tab]
\CodeBefore
\Body
\Block{1-7}{$T^{ \uparrow, +1}_{322}$} \\
  $x_{4}$
& $x_{5}$
& $x_{6}$
& $s_{7}$
& $ \vec{k}_{7 \perp } $
& $s_{8}$
& $x_{8}$
\\
  $- \dfrac{x_{1}}{1 - \xi } $
& $- \dfrac{x_{2}}{1 - \xi } $
& $ \dfrac{1 + \xi }{1 - \xi } - \dfrac{x_{3}}{1 - \xi } $
& $+ \dfrac{1}{2} $
& $ \vec{k}_{6 \perp } $
& $+ \dfrac{1}{2} $
& $ \dfrac{x_{3}}{1 + \xi } $
\\
  $- \dfrac{x_{1}}{1 - \xi } $
& $- \dfrac{x_{2}}{1 - \xi } $
& $ \dfrac{1 + \xi }{1 - \xi } - \dfrac{x_{3}}{1 - \xi } $
& $- \dfrac{1}{2} $
& $ \vec{k}_{6 \perp } $
& $+ \dfrac{1}{2} $
& $ \dfrac{x_{3}}{1 + \xi } $
\\
  $ \dfrac{1 + \xi }{1 - \xi } - \dfrac{x_{1}}{1 - \xi } $
& $- \dfrac{x_{2}}{1 - \xi } $
& $- \dfrac{x_{3}}{1 - \xi } $
& $+ \dfrac{1}{2} $
& $ \vec{k}_{4 \perp } $
& $- \dfrac{1}{2} $
& $ \dfrac{x_{1}}{1 + \xi } $
\\
  $- \dfrac{x_{1}}{1 - \xi } $
& $ \dfrac{1 + \xi }{1 - \xi } - \dfrac{x_{2}}{1 - \xi } $
& $- \dfrac{x_{3}}{1 - \xi } $
& $- \dfrac{1}{2} $
& $ \vec{k}_{5 \perp } $
& $+ \dfrac{1}{2} $
& $ \dfrac{x_{2}}{1 + \xi } $
\\
  $ \dfrac{1 + \xi }{1 - \xi } - \dfrac{x_{2}}{1 - \xi } $
& $- \dfrac{x_{3}}{1 - \xi } $
& $- \dfrac{x_{1}}{1 - \xi } $
& $+ \dfrac{1}{2} $
& $ \vec{k}_{4 \perp } $
& $+ \dfrac{1}{2} $
& $ \dfrac{x_{2}}{1 + \xi } $
\\
  $- \dfrac{x_{1}}{1 - \xi } $
& $- \dfrac{x_{3}}{1 - \xi } $
& $ \dfrac{1 + \xi }{1 - \xi } - \dfrac{x_{2}}{1 - \xi } $
& $+ \dfrac{1}{2} $
& $ \vec{k}_{6 \perp } $
& $+ \dfrac{1}{2} $
& $ \dfrac{x_{2}}{1 + \xi } $
\\
\end{NiceTabular}
\begin{tikzpicture}[remember picture, overlay]
\coordinate (a) at (t322tab-row-5-|t322tab-col-1);
\coordinate (b) at (t322tab-row-5-|t322tab-col-8);
\draw[line width = 3pt] (a) -- (b);
\end{tikzpicture}
\caption{Constrained variables for the nonzero terms of $T^{ \uparrow, +1}_{322}$. Contributions from the initial photon splitting into $d \overline{d} $ and into $u \overline{u} $ are above and below the dark  line, respectively.}
\label{tthreetwotwo_tab}
\end{table}
\vspace*{\fill}
\begin{table}[H]
\centering
\begin{NiceTabular}{ccccccc}[hvlines, name = t232tab]
\CodeBefore
\Body
\Block{1-7}{$T^{ \uparrow, +1}_{232}$} \\
  $x_{4}$
& $x_{5}$
& $x_{6}$
& $s_{7}$
& $ \vec{k}_{7 \perp } $
& $s_{8}$
& $x_{8}$
\\
  $- \dfrac{x_{2}}{1 - \xi } $
& $- \dfrac{x_{1}}{1 - \xi } $
& $ \dfrac{1 + \xi }{1 - \xi } - \dfrac{x_{3}}{1 - \xi } $
& $+ \dfrac{1}{2} $
& $ \vec{k}_{6 \perp } $
& $+ \dfrac{1}{2} $
& $ \dfrac{x_{3}}{1 + \xi } $
\\
  $- \dfrac{x_{2}}{1 - \xi } $
& $- \dfrac{x_{1}}{1 - \xi } $
& $ \dfrac{1 + \xi }{1 - \xi } - \dfrac{x_{3}}{1 - \xi } $
& $- \dfrac{1}{2} $
& $ \vec{k}_{6 \perp } $
& $+ \dfrac{1}{2} $
& $ \dfrac{x_{3}}{1 + \xi } $
\\
  $ \dfrac{1 + \xi }{1 - \xi } - \dfrac{x_{2}}{1 - \xi } $
& $- \dfrac{x_{1}}{1 - \xi } $
& $- \dfrac{x_{3}}{1 - \xi } $
& $+ \dfrac{1}{2} $
& $ \vec{k}_{4 \perp } $
& $- \dfrac{1}{2} $
& $ \dfrac{x_{2}}{1 + \xi } $
\\
  $- \dfrac{x_{2}}{1 - \xi } $
& $ \dfrac{1 + \xi }{1 - \xi } - \dfrac{x_{1}}{1 - \xi } $
& $- \dfrac{x_{3}}{1 - \xi } $
& $- \dfrac{1}{2} $
& $ \vec{k}_{5 \perp } $
& $+ \dfrac{1}{2} $
& $ \dfrac{x_{1}}{1 + \xi } $
\\
  $ \dfrac{1 + \xi }{1 - \xi } - \dfrac{x_{1}}{1 - \xi } $
& $- \dfrac{x_{3}}{1 - \xi } $
& $- \dfrac{x_{2}}{1 - \xi } $
& $+ \dfrac{1}{2} $
& $ \vec{k}_{4 \perp } $
& $+ \dfrac{1}{2} $
& $ \dfrac{x_{1}}{1 + \xi } $
\\
  $- \dfrac{x_{2}}{1 - \xi } $
& $- \dfrac{x_{3}}{1 - \xi } $
& $ \dfrac{1 + \xi }{1 - \xi } - \dfrac{x_{1}}{1 - \xi } $
& $+ \dfrac{1}{2} $
& $ \vec{k}_{6 \perp } $
& $+ \dfrac{1}{2} $
& $ \dfrac{x_{1}}{1 + \xi } $
\\
\end{NiceTabular}
\begin{tikzpicture}[remember picture, overlay]
\coordinate (a) at (t232tab-row-5-|t232tab-col-1);
\coordinate (b) at (t232tab-row-5-|t232tab-col-8);
\draw[line width = 3pt] (a) -- (b);
\end{tikzpicture}
\caption{Constrained variables for the nonzero terms of $T^{ \uparrow, +1}_{232}$. Contributions from the initial photon splitting into $d \overline{d} $ and into $u \overline{u} $ are above and below the dark  line, respectively.}
\label{ttwothreetwo_tab}
\end{table}
\vspace*{\fill}
\newpage
\begin{table}[H]
\centering
\begin{NiceTabular}{ccccccc}[hvlines, name = t223tab]
\CodeBefore
\Body
\Block{1-7}{ $T^{ \uparrow, +1}_{223} $ } \\
  $x_{4}$
& $x_{5}$
& $x_{6}$
& $s_{7}$
& $ \vec{k}_{7 \perp } $
& $s_{8}$
& $x_{8}$
\\
  $ \dfrac{1 + \xi }{1 - \xi } - \dfrac{x_{3}}{1 - \xi } $
& $- \dfrac{x_{1}}{1 - \xi } $
& $- \dfrac{x_{2}}{1 - \xi } $
& $+ \dfrac{1}{2} $
& $ \vec{k}_{4 \perp } $
& $- \dfrac{1}{2} $
& $ \dfrac{x_{3}}{1 + \xi } $
\\
  $ \dfrac{1 + \xi }{1 - \xi } - \dfrac{x_{3}}{1 - \xi } $
& $- \dfrac{x_{2}}{1 - \xi } $
& $- \dfrac{x_{1}}{1 - \xi } $
& $+ \dfrac{1}{2} $
& $ \vec{k}_{4 \perp } $
& $- \dfrac{1}{2} $
& $ \dfrac{x_{3}}{1 + \xi } $
\\
  $ \dfrac{1 + \xi }{1 - \xi } - \dfrac{x_{1}}{1 - \xi } $
& $- \dfrac{x_{2}}{1 - \xi } $
& $- \dfrac{x_{3}}{1 - \xi } $
& $+ \dfrac{1}{2} $
& $ \vec{k}_{4 \perp } $
& $+ \dfrac{1}{2} $
& $ \dfrac{x_{1}}{1 + \xi } $
\\
  $ \dfrac{1 + \xi }{1 - \xi } - \dfrac{x_{2}}{1 - \xi } $
& $- \dfrac{x_{1}}{1 - \xi } $
& $- \dfrac{x_{3}}{1 - \xi } $
& $+ \dfrac{1}{2} $
& $ \vec{k}_{4 \perp } $
& $+ \dfrac{1}{2} $
& $ \dfrac{x_{2}}{1 + \xi } $
\\
  $- \dfrac{x_{3}}{1 - \xi } $
& $ \dfrac{1 + \xi }{1 - \xi } - \dfrac{x_{1}}{1 - \xi } $
& $- \dfrac{x_{2}}{1 - \xi } $
& $- \dfrac{1}{2} $
& $ \vec{k}_{5 \perp } $
& $+ \dfrac{1}{2} $
& $ \dfrac{x_{1}}{1 + \xi } $
\\
  $- \dfrac{x_{3}}{1 - \xi } $
& $ \dfrac{1 + \xi }{1 - \xi } - \dfrac{x_{2}}{1 - \xi } $
& $- \dfrac{x_{1}}{1 - \xi } $
& $- \dfrac{1}{2} $
& $ \vec{k}_{5 \perp } $
& $+ \dfrac{1}{2} $
& $ \dfrac{x_{2}}{1 + \xi } $
\\
  $- \dfrac{x_{3}}{1 - \xi } $
& $- \dfrac{x_{2}}{1 - \xi } $
& $ \dfrac{1 + \xi }{1 - \xi } - \dfrac{x_{1}}{1 - \xi } $
& $- \dfrac{1}{2} $
& $ \vec{k}_{6 \perp } $
& $+ \dfrac{1}{2} $
& $ \dfrac{x_{1}}{1 + \xi } $
\\
  $- \dfrac{x_{3}}{1 - \xi } $
& $- \dfrac{x_{1}}{1 - \xi } $
& $ \dfrac{1 + \xi }{1 - \xi } - \dfrac{x_{2}}{1 - \xi } $
& $- \dfrac{1}{2} $
& $ \vec{k}_{6 \perp } $
& $+ \dfrac{1}{2} $
& $ \dfrac{x_{2}}{1 + \xi } $
\\
\end{NiceTabular}
\begin{tikzpicture}[remember picture, overlay]
\coordinate (a) at (t223tab-row-5-|t223tab-col-1);
\coordinate (b) at (t223tab-row-5-|t223tab-col-8);
\draw[line width = 3pt] (a) -- (b);
\end{tikzpicture}
\caption{Constrained variables for the nonzero terms of $T^{ \uparrow, +1}_{223}$. Contributions from the initial photon splitting into $d \overline{d} $ and into $u \overline{u} $ are above and below the dark  line, respectively.}
\label{ttwotwothree_tab}
\end{table}
\begin{table}[H]
\centering
\begin{NiceTabular}{ccccccc}[hvlines, name = t333tab]
\CodeBefore
\Body
\Block{1-7}{ $T^{ \uparrow, -1}_{333}  \left( \gamma \rightarrow d \overline{d} \right) $ } \\
  $x_{4}$
& $x_{5}$
& $x_{6}$
& $s_{7}$
& $ \overrightarrow{k}_{7 \perp } $
& $s_{8}$
& $x_{8}$
\\
  $- \dfrac{x_{1}}{1 - \xi } $
& $- \dfrac{x_{2}}{1 - \xi } $
& $ \dfrac{1 + \xi }{1 - \xi } - \dfrac{x_{3}}{1 - \xi } $
& $+ \dfrac{1}{2} $
& $ \overrightarrow{k}_{6 \perp } $
& $- \dfrac{1}{2} $
& $ \dfrac{x_{3}}{1 + \xi } $
\\
  $- \dfrac{x_{2}}{1 - \xi } $
& $- \dfrac{x_{1}}{1 - \xi } $
& $ \dfrac{1 + \xi }{1 - \xi } - \dfrac{x_{3}}{1 - \xi } $
& $+ \dfrac{1}{2} $
& $ \overrightarrow{k}_{6 \perp } $
& $- \dfrac{1}{2} $
& $ \dfrac{x_{3}}{1 + \xi } $
\\
  $- \dfrac{x_{1}}{1 - \xi } $
& $ \dfrac{1 + \xi }{1 - \xi } - \dfrac{x_{3}}{1 - \xi } $
& $- \dfrac{x_{2}}{1 - \xi } $
& $- \dfrac{1}{2} $
& $ \overrightarrow{k}_{5 \perp } $
& $- \dfrac{1}{2} $
& $ \dfrac{x_{3}}{1 + \xi } $
\\
  $- \dfrac{x_{2}}{1 - \xi } $
& $ \dfrac{1 + \xi }{1 - \xi } - \dfrac{x_{3}}{1 - \xi } $
& $- \dfrac{x_{1}}{1 - \xi } $
& $- \dfrac{1}{2} $
& $ \overrightarrow{k}_{5 \perp } $
& $- \dfrac{1}{2} $
& $ \dfrac{x_{3}}{1 + \xi } $
\\
  $- \dfrac{x_{1}}{1 - \xi } $
& $ \dfrac{1 + \xi }{1 - \xi } - \dfrac{x_{3}}{1 - \xi } $
& $- \dfrac{x_{2}}{1 - \xi } $
& $+ \dfrac{1}{2} $
& $ \overrightarrow{k}_{5 \perp } $
& $- \dfrac{1}{2} $
& $ \dfrac{x_{3}}{1 + \xi } $
\\
  $- \dfrac{x_{2}}{1 - \xi } $
& $ \dfrac{1 + \xi }{1 - \xi } - \dfrac{x_{3}}{1 - \xi } $
& $- \dfrac{x_{1}}{1 - \xi } $
& $+ \dfrac{1}{2} $
& $ \overrightarrow{k}_{5 \perp } $
& $- \dfrac{1}{2} $
& $ \dfrac{x_{3}}{1 + \xi } $
\\
\end{NiceTabular}
\caption{Constrained variables for the nonzero terms of $T^{ \uparrow, -1}_{333}$ from the initial photon splitting into $d \overline{d} $.}
\label{tthreethreethree_tab_partone}
\end{table}
\begin{table}[H]
\centering
\begin{NiceTabular}{ccccccc}[hvlines]
\CodeBefore
\Body
\Block{1-7}{ $T^{ \uparrow, -1}_{333}  \left( \gamma \rightarrow u \overline{u} \right) $ } \\
  $x_{4}$
& $x_{5}$
& $x_{6}$
& $s_{7}$
& $ \overrightarrow{k}_{7 \perp } $
& $s_{8}$
& $x_{8}$
\\
 $ \dfrac{1 + \xi }{1 - \xi } - \dfrac{x_{1}}{1 - \xi } $
& $- \dfrac{x_{2}}{1 - \xi } $
& $- \dfrac{x_{3}}{1 - \xi } $
& $+ \dfrac{1}{2} $
& $ \overrightarrow{k}_{4 \perp } $
& $- \dfrac{1}{2} $
& $ \dfrac{x_{1}}{1 + \xi } $
\\
  $ \dfrac{1 + \xi }{1 - \xi } - \dfrac{x_{2}}{1 - \xi } $
& $- \dfrac{x_{1}}{1 - \xi } $
& $- \dfrac{x_{3}}{1 - \xi } $
& $+ \dfrac{1}{2} $
& $ \overrightarrow{k}_{4 \perp } $
& $- \dfrac{1}{2} $
& $ \dfrac{x_{2}}{1 + \xi } $
\\
  $- \dfrac{x_{2}}{1 - \xi } $
& $ \dfrac{1 + \xi }{1 - \xi } - \dfrac{x_{1}}{1 - \xi } $
& $- \dfrac{x_{3}}{1 - \xi } $
& $- \dfrac{1}{2} $
& $ \overrightarrow{k}_{5 \perp } $
& $- \dfrac{1}{2} $
& $ \dfrac{x_{1}}{1 + \xi } $
\\
  $- \dfrac{x_{1}}{1 - \xi } $
& $ \dfrac{1 + \xi }{1 - \xi } - \dfrac{x_{2}}{1 - \xi } $
& $- \dfrac{x_{3}}{1 - \xi } $
& $- \dfrac{1}{2} $
& $ \overrightarrow{k}_{5 \perp } $
& $- \dfrac{1}{2} $
& $ \dfrac{x_{2}}{1 + \xi } $
\\
  $- \dfrac{x_{2}}{1 - \xi } $
& $ \dfrac{1 + \xi }{1 - \xi } - \dfrac{x_{1}}{1 - \xi } $
& $- \dfrac{x_{3}}{1 - \xi } $
& $+ \dfrac{1}{2} $
& $ \overrightarrow{k}_{5 \perp } $
& $- \dfrac{1}{2} $
& $ \dfrac{x_{1}}{1 + \xi } $
\\
  $- \dfrac{x_{1}}{1 - \xi } $
& $ \dfrac{1 + \xi }{1 - \xi } - \dfrac{x_{2}}{1 - \xi } $
& $- \dfrac{x_{3}}{1 - \xi } $
& $+ \dfrac{1}{2} $
& $ \overrightarrow{k}_{5 \perp } $
& $- \dfrac{1}{2} $
& $ \dfrac{x_{2}}{1 + \xi } $
\\
  $ \dfrac{1 + \xi }{1 - \xi } - \dfrac{x_{1}}{1 - \xi } $
& $- \dfrac{x_{3}}{1 - \xi } $
& $- \dfrac{x_{2}}{1 - \xi } $
& $+ \dfrac{1}{2} $
& $ \overrightarrow{k}_{4 \perp } $
& $- \dfrac{1}{2} $
& $ \dfrac{x_{1}}{1 + \xi } $
\\
  $ \dfrac{1 + \xi }{1 - \xi } - \dfrac{x_{2}}{1 - \xi } $
& $- \dfrac{x_{3}}{1 - \xi } $
& $- \dfrac{x_{1}}{1 - \xi } $
& $+ \dfrac{1}{2} $
& $ \overrightarrow{k}_{4 \perp } $
& $- \dfrac{1}{2} $
& $ \dfrac{x_{2}}{1 + \xi } $
\\
  $- \dfrac{x_{2}}{1 - \xi } $
& $- \dfrac{x_{3}}{1 - \xi } $
& $ \dfrac{1 + \xi }{1 - \xi } - \dfrac{x_{1}}{1 - \xi } $
& $+ \dfrac{1}{2} $
& $ \overrightarrow{k}_{6 \perp } $
& $- \dfrac{1}{2} $
& $ \dfrac{x_{1}}{1 + \xi } $
\\
  $- \dfrac{x_{1}}{1 - \xi } $
& $- \dfrac{x_{3}}{1 - \xi } $
& $ \dfrac{1 + \xi }{1 - \xi } - \dfrac{x_{2}}{1 - \xi } $
& $+ \dfrac{1}{2} $
& $ \overrightarrow{k}_{6 \perp } $
& $- \dfrac{1}{2} $
& $ \dfrac{x_{2}}{1 + \xi } $
\\
\end{NiceTabular}
\caption{Constrained variables for the nonzero terms of $T^{ \uparrow, -1}_{333}$ from the initial photon splitting into $u \overline{u} $.}
\label{tthreethreethree_tab_partwo}
\end{table}

\subsection[Numerical Results on the Transition Distribution Amplitudes]{Numerical Results}
\label{subsec_numerical_results}

\begin{figure}[h]
\centering
\subfloat[]{\includegraphics[width = 0.47\textwidth]{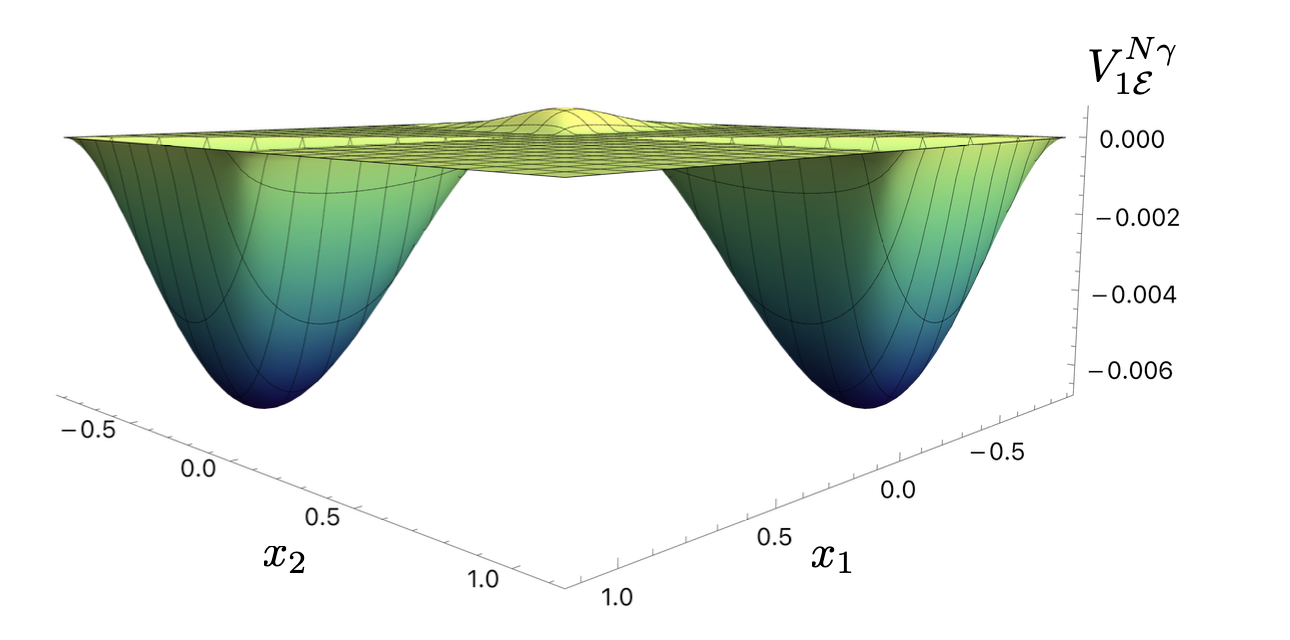}
\label{v1e_plot}}\quad\quad
\subfloat[]{\includegraphics[width = 0.47\textwidth]{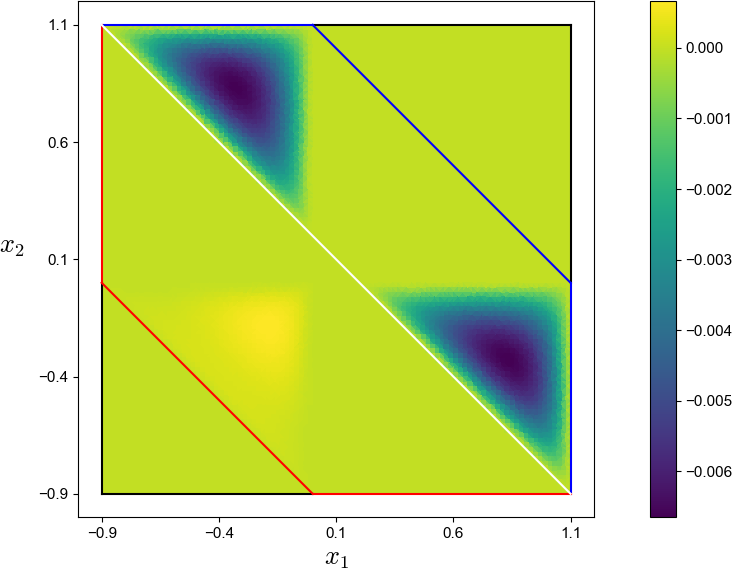}
\label{v1e_densityplot}}
\caption{(Color online.) Results for the photon-to-proton $V_{1 \mathcal{E}}$ TDA for $ \xi = 0.1$. (a) 3D plot as a function of $x_{1}, x_{2}$. (b) Density plot as a function of $x_{1}, x_{2}$. On the (white) diagonal, $x_{3} = 0$. Inside the upper trapezoid, $ \xi - 1 < x_{3} < 0$, with $x_{3} = \xi - 1$ on the upper slanted (blue) line. Inside the lower trapezoid, $0 < x_{3} < \xi + 1$, with $x_{3} = \xi + 1$ on the lower slanted (red) line. }
\label{fig_v1e}
\end{figure}

\begin{figure}[h]
\centering
\subfloat[]{\includegraphics[width = 0.47\textwidth]{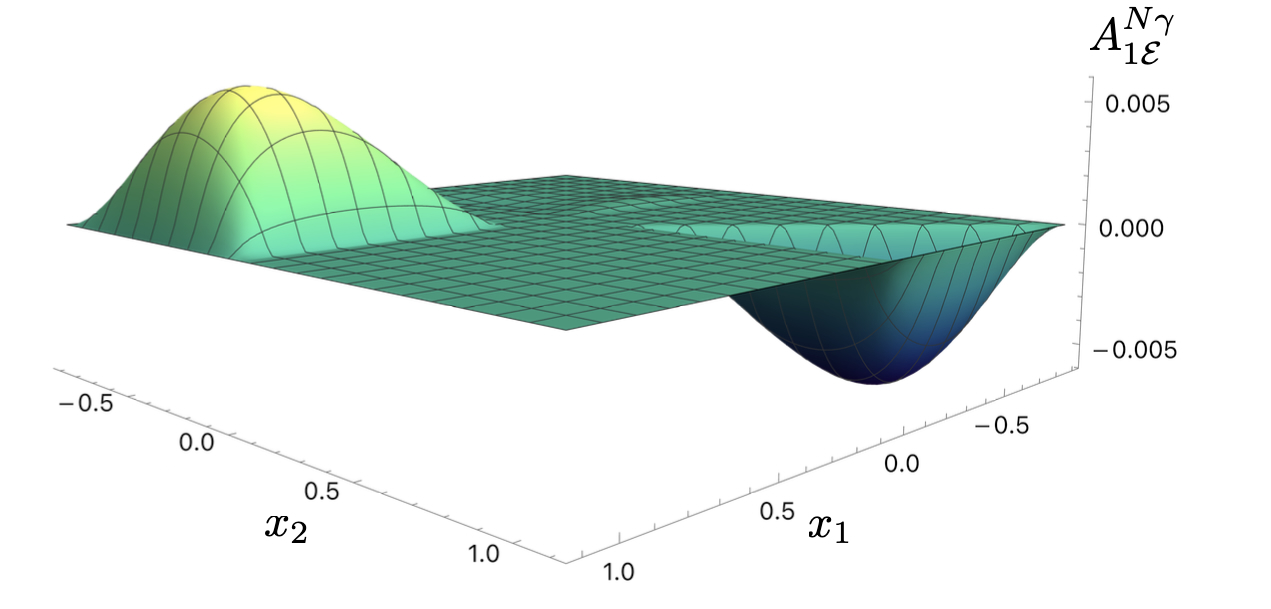}
\label{a1e_plot}}\quad\quad
\subfloat[]{\includegraphics[width = 0.47\textwidth]{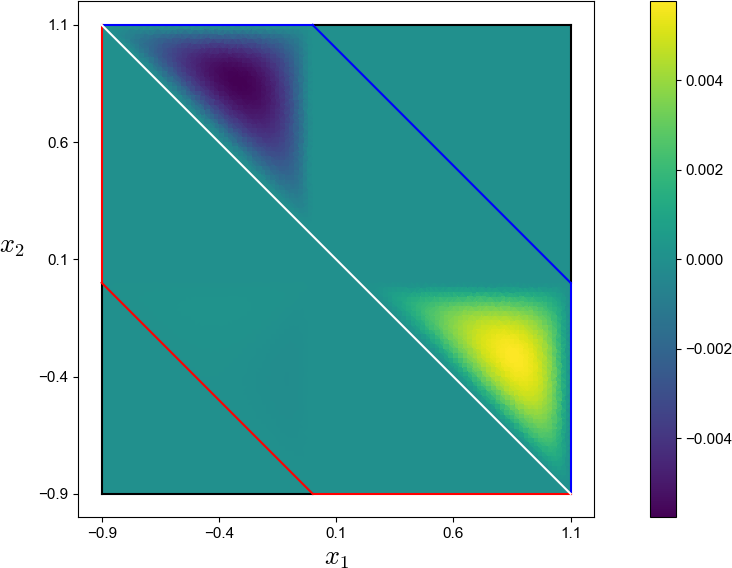}
\label{a1e_densityplot}}
\caption{The same as in Fig.~\ref{fig_v1e} for the photon-to-proton $A_{1 \mathcal{E}}$ TDA for $ \xi = 0.1$.}
\label{fig_a1e}
\end{figure}

\begin{figure}[h]
\centering
\subfloat[]{\includegraphics[width = 0.47\textwidth]{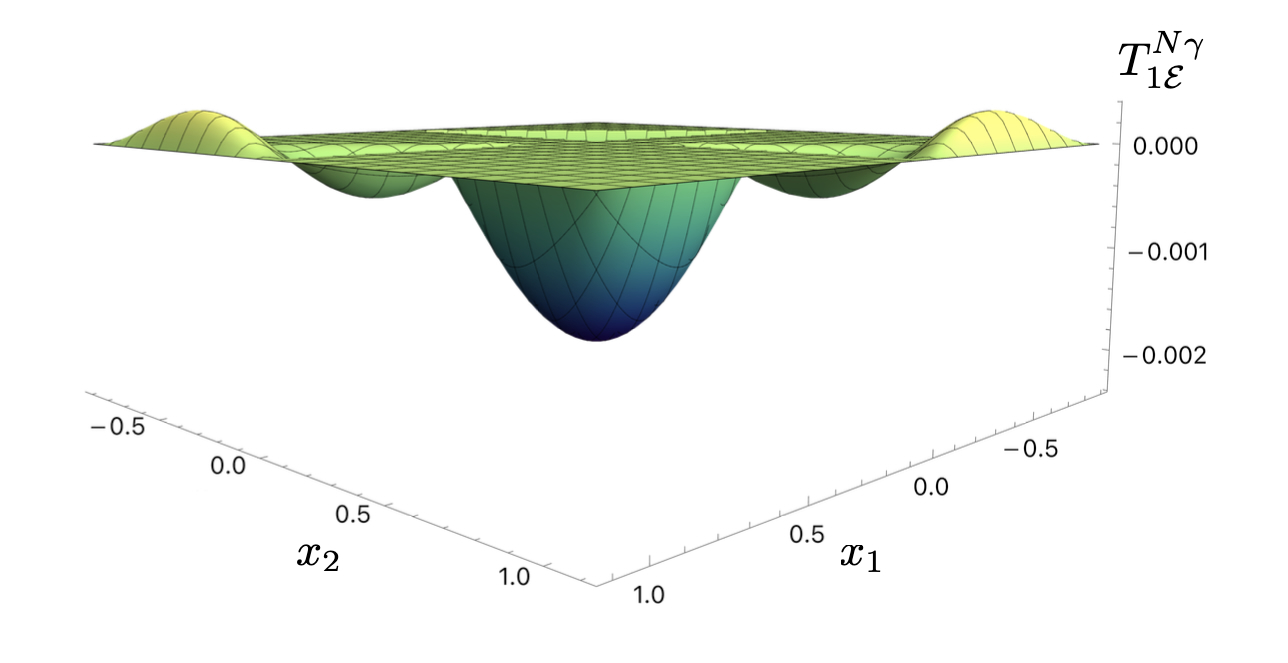}
\label{t1e_plot}}\quad\quad
\subfloat[]{\includegraphics[width = 0.47\textwidth]{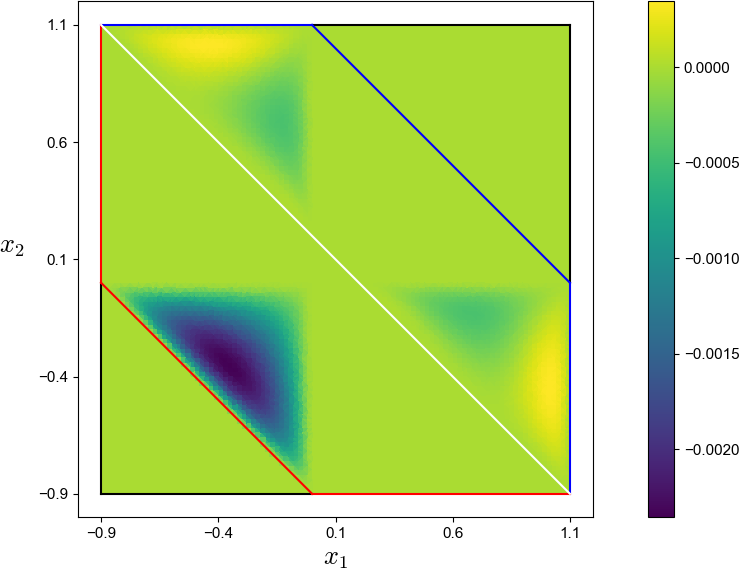}
\label{t1e_densityplot}}
\caption{The same as in Fig.~\ref{fig_v1e} for the photon-to-proton $T_{1 \mathcal{E}}$ TDA for $ \xi = 0.1$.}
\label{fig_t1e}
\end{figure}

\begin{figure}[h]
\centering
\subfloat[]{\includegraphics[width = 0.47\textwidth]{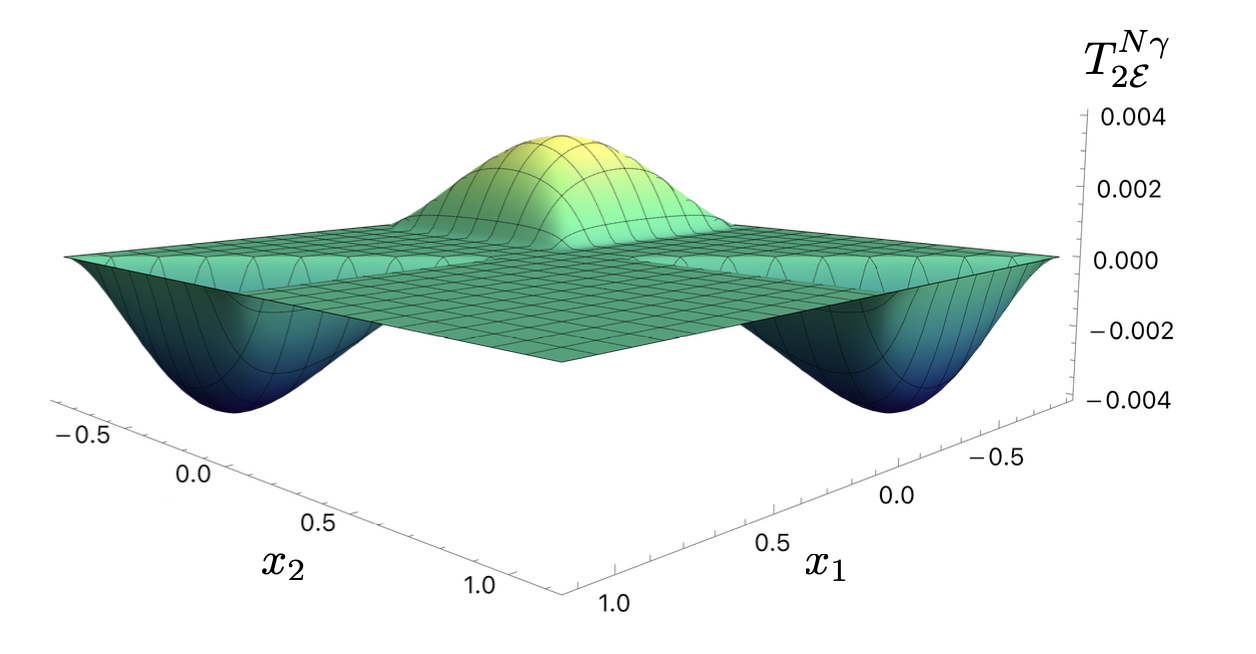}
\label{t2e_plot}} \quad\quad
\subfloat[]{\includegraphics[width = 0.47\textwidth]{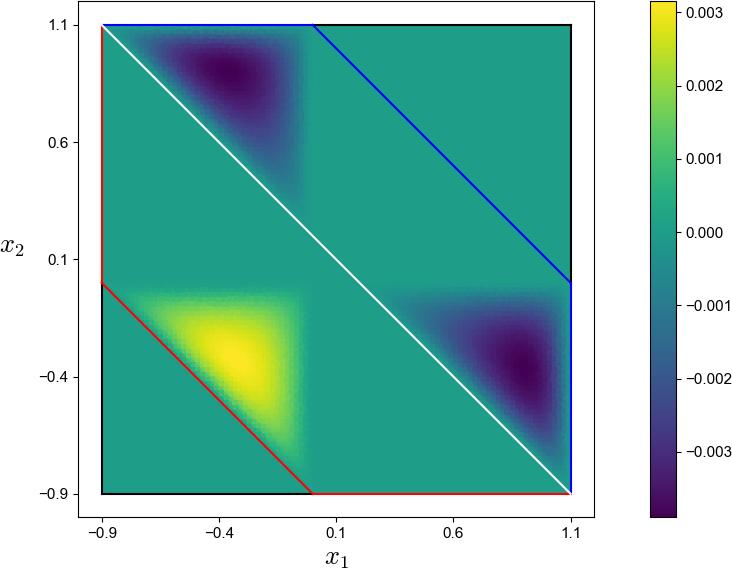}
\label{t2e_densityplot}}
\caption{The same as in Fig.~\ref{fig_v1e} for the photon-to-proton $T_{2 \mathcal{E}}$ TDA for $ \xi = 0.1$.}
\label{fig_t2e}
\end{figure}

In this section, we present numerical predictions for the photon-to-proton TDAs within the light-front model presented in Sec.~\ref{subsec_lfd_model}. We work in the backward kinematical region, corresponding to transverse momentum $ \Delta_{T} $ of the proton with respect to the photon equal to zero, and to values of the skewness variable $ \xi $ close to zero. As an example, we chose $ \xi = 0.1 $, which, by Eq.~\eqref{sudakov_u}, corresponds to $ \lvert u \rvert = 0.196 \, \text{GeV}^{2} $. The results are shown in Figs.~\ref{fig_v1e}--\ref{fig_t2e}. Making use of the constraint~\eqref{delta_of_extracted_pmomemntum}, we  present 3D plots for the TDAs as functions of $x_1$ and $x_2$ in Figs.~\ref{v1e_plot}--\ref{t2e_plot}, while Figs.~\ref{v1e_densityplot}--\ref{t2e_densityplot} are the corresponding density plots. The density plots have been divided in various regions according to the value of $x_{3}$. On the (white) diagonal, we have $x_{3} = 0$, and the momentum fraction is positive below the line and negative above the line. It decreases moving towards the upper slanted (blue) line, where it reaches its minimal value of $ \xi - 1$, while it increases towards the lower slanted (red) line, where we have the maximum value of $ \xi + 1$. Since we are truncating the Fock expansion of the initial photon to the leading light-quark--antiquark-pair component, the soft transition into the final nucleon is schematically represented in Fig.~\ref{dglapone_fig}. The extracted antiquark corresponds to positive values of one of the variables $x_{1}, x_{2}, x_{3}$, while the absorbed quarks to negative values of the remaining two. We can check from the figures that the model has the correct support. Furthermore, $V_{1 \mathcal{E} }, T_{1 \mathcal{E} }, T_{2 \mathcal{E} }$ are symmetric under the exchange of $x_{1}$ and $x_{2}$, while $A_{1 \mathcal{E} }$ is antisymmetric, as expected from Eqs.~\eqref{bigv_one}--\eqref{bigt_two}. In the case of $V_{1 \mathcal{E} }, A_{1 \mathcal{E} }$, this is true even in the support region where both variables are negative, where an antiquark down is extracted from the photon, even though it is difficult to see, since the contribution is highly suppressed.
\newpage
\vspace*{\fill}
\vspace*{\fill}
\vspace*{\fill}
\newpage

The Mellin moments of the TDAs are defined as
\begin{equation}
S^{ \left( a, b, c \right) } = \int \! dx_{1}dx_{2}dx_{3} \delta \! \left( x_{1} + x_{2} + x_{3} - 2 \xi \right) x_{1}^{a}x_{2}^{b}x_{3}^{c} S \! \left( x_{1}, x_{2}, x_{3} \right) \! ,
\label{mellin_mom}
\end{equation}
where $S$ is one of the TDAs. To give a more quantitative representation of the backward photon-to-proton TDAs, we evaluated the $(0, 0, 0)$-moments $ S^{ \left( 0, 0, 0 \right) }$. The results are given in Tab.~\ref{tdas_mellin_mom_trizero}, alongside the separate contributions from every possible orbital-angular-momentum component of the three-quark system in the proton. The non-admissible terms are denoted with a slash. We already know that the integral of the antisymmetric $A_{1 \mathcal{E} }$ should be exactly zero, while it turns out to be negligible for $L_{z} = +1$ in $T_{1 \mathcal{E} }, T_{2 \mathcal{E} }$, which contributes only to the helicity amplitude $T^{ \uparrow, +1}_{223}$.

\section{Conclusions}
\label{section_conclusions}

A factorization of the scattering amplitude of TCS in the backward kinematical region has been proposed in Ref.~\cite{Pire:2022fbi}. Analogously to other backward exclusive reactions, the low-energy subprocess that splits the initial nucleon into its parton content is represented by nucleon DAs, while the soft transition from the initial real photon to the final nucleon is encoded in photon-to-nucleon TDAs. A natural framework to study these non-perturbative objects is LFD, where the interacting states can be expanded in the Fock space, and a clear partonic interpretation emerges. Nucleon distribution amplitudes have already been investigated taking advantage of these techniques in Ref.~\cite{Pasquini:2009ki}. This work follows in those steps, aiming for a phenomenologically solid model of the photon-to-nucleon TDAs. Currently, there is no formal proof of the factorization for backward TCS, but the hypothesis is supported by the analogy with the well-established case of forward scattering. The analysis of existing data from Jefferson Lab is also in agreement with the outset of factorization through TDAs in the backward region of exclusive reactions involving mesons~\cite{CLAS:2017rgp, JeffersonLabFp:2019gpp, CLAS:2020yqf}.

The study of backward exclusive reactions is a promising subject, hence the request for effective models, also for the planning of experiments at the, currently under construction, Electron-Ion Collider~\cite{AbdulKhalek:2021gbh}. This work begins the endeavour for modeling the  photon-to-nucleon TDAs, focusing on the leading contribution where two light quarks take the place of a light antiquark in the parton configuration of the initial photon, in order to make up the final nucleon. The results exhibit the basic features that are to be expected, and the first Mellin moments have been numerically evaluated, separating the individual contribution from the possible orbital-angular-momentum components of the proton LFWFs. In the future, the same techniques could be used to investigate the other fundamental ways in which the transition can happen, involving higher-order Fock states of the photon and the proton. The scale of the model, about $0.5 \, \text{GeV} $, is dictated by the effective description of the nucleon as three constituent valence quarks. Understanding the scale evolution will be key to compare theoretical predictions against the experimental data expected for the near future and work in this direction is in progress.

\begin{table}[ht]
\centering
\begin{NiceTabular}{cWc{1.3cm}Wc{1.3cm}Wc{1.3cm}Wc{1.3cm}Wc{1.3cm}c}[corners, hvlines]
\CodeBefore
\Body

& Total
& $L_{z} = 0$
& $L_{z} = +1$
& $L_{z} = -1$
& $L_{z} = +2$
&
\\
  $ V_{1 \mathcal{E} }^{ \left( 0, 0, 0 \right) } $
& $-2.3$
& $-1.0$
& $-1.3$
& /
& /
& \Block{4-1}{$ \times 10^{-3}$}
\\
  $ A_{1 \mathcal{E} }^{ \left( 0, 0, 0 \right) } $
& 0
& 0
& 0
& /
& /
\\
  $ T_{1 \mathcal{E} }^{ \left( 0, 0, 0 \right) } $
& $-0.4$
& $-0.2$
& $ \sim 0 $
& $-0.2$
& /
\\
  $ T_{2 \mathcal{E} }^{ \left( 0, 0, 0 \right) } $
& $-0.8$
& $-1.0$
& $ \sim 0 $
& $+0.2$
& /
\\
\end{NiceTabular}
\caption{Mellin moments $ \left( 0, 0, 0 \right) $ of photon-to-proton TDAs for $ \xi = 0.1$. The total results are shown in the second column, while columns 3--6 show the results from the individual partial waves of the proton LFWF. The entries with a slash are forbidden by angular momentum conservation.}
\label{tdas_mellin_mom_trizero}
\end{table}
%
%
%

\acknowledgments

We thank B. Pire, K. Semenov, and L. Szymanowski for useful discussions and for their continuous interest and motivation in pursuing this work.

\begin{appendix}

\section{Conventions for Light-Front Dynamics}
\label{appendix_lfd}

The Dirac gamma matrices in the chiral representation are
\begin{equation}
\gamma^{0} = \begin{pmatrix}
0_{2 \times 2} & I_{2 \times 2} \\
I_{2 \times 2} & 0_{2 \times 2}
\end{pmatrix} \! , \qquad \gamma^{j} = \begin{pmatrix}
0_{2 \times 2}    & -\sigma^{j} \\
\sigma^{j} & 0_{2 \times 2}
\end{pmatrix} \! , \qquad \gamma^{5} = \begin{pmatrix}
I_{2 \times 2} & 0_{2 \times 2}  \\
0_{2 \times 2} & -I_{2 \times 2}
\end{pmatrix} \! ,
\label{ks_chiral_gammatrices}
\end{equation}
where $j = 1, 2, 3$, $I_{2 \times 2}$ is the 2$ \times $2 identity matrix, $0_{2 \times 2}$ is the 2$ \times $2 matrix of all zeros, and $ \sigma^{j} $ are the usual Pauli matrices. The charge conjugation matrix is
\begin{equation}
\mathcal{C} = i \gamma^{2} \gamma^{0} = \begin{pmatrix}
-i \sigma^{2} & 0           \\
0             & i \sigma^{2}
\end{pmatrix} \! ,
\label{ks_chiral_charge_conjugation_matrix}
\end{equation}
and the change of basis matrix to the usual Dirac representation is
\begin{equation}
U = \frac{1}{ \sqrt{2} } \begin{pmatrix}
I_{2 \times 2} & I_{2 \times 2}  \\
I_{2 \times 2} & -I_{2 \times 2}
\end{pmatrix} \! .
\label{changeofbasis_diractoks}
\end{equation}

We define the following operators on Dirac-spinor space:
\begin{equation}
\Lambda_{\pm} = \frac{1}{2} \gamma^{0} \gamma^{\pm},
\label{plusminus_projectors}
\end{equation}
which have all the properties of a complete set of orthogonal projectors. The projections with $ \Lambda_{+} $ are called good or large components, and the projections with $ \Lambda_{-} $ are called bad or small components. In the chiral 
representation~\eqref{ks_chiral_gammatrices}, they are already in diagonal form:
\begin{equation}
\Lambda_{+} = \begin{pmatrix}
1 & 0 & 0 & 0 \\
0 & 0 & 0 & 0 \\
0 & 0 & 0 & 0 \\
0 & 0 & 0 & 1
\end{pmatrix} \! , \qquad
\Lambda_{-} = \begin{pmatrix}
0 & 0 & 0 & 0 \\
0 & 1 & 0 & 0 \\
0 & 0 & 1 & 0 \\
0 & 0 & 0 & 0
\end{pmatrix} \! .
\label{kogsopbasis_pm}
\end{equation}
The LC helicity spinors are
\begin{alignat}{2}
u^{{\rm LC}}_{ \uparrow } \! \left( p \right)
& = \frac{1}{ \sqrt{p^{+}} } \begin{pmatrix}
p^{+}          \\
p^{1} + ip^{2} \\
m              \\
0
\end{pmatrix} \! , \qquad
& u^{{\rm LC}}_{ \downarrow } \! \left( p \right)
& = \frac{1}{ \sqrt{p^{+}} } \begin{pmatrix}
0               \\
m               \\
-p^{1} + ip^{2} \\
p^{+}
\end{pmatrix} \! ,
\label{lc_us} \\
v^{{\rm LC}}_{ \uparrow } \! \left( p \right)
& = \frac{1}{ \sqrt{p^{+}} } \begin{pmatrix}
0               \\
-m              \\
-p^{1} + ip^{2} \\
p^{+}
\end{pmatrix} \! , \qquad
& v^{{\rm LC}}_{ \downarrow } \! \left( p \right)
& = \frac{1}{ \sqrt{p^{+}} }
 \begin{pmatrix}
p^{+}           \\
p^{1} + ip^{2}  \\
-m              \\
0
\end{pmatrix} \! .
\label{lc_vs}
\end{alignat}
Therefore, the Dirac adjoints of the good components of the LC helicity spinors are
\begin{alignat}{2}
\overline{u}_{ \uparrow } \! \left( p \right)
& = \frac{1}{ \sqrt{p^{+}} } 
\begin{pmatrix}
0 & 0 & p^{+} & 0 \end{pmatrix} =\overline{v}_{ \downarrow } \! \left( p \right) \, ,  \qquad
& \overline{u}_{ \downarrow } \! \left( p \right)
& = \frac{1}{ \sqrt{p^{+}} } 
\begin{pmatrix}
0 & p^{+} & 0 & 0 \end{pmatrix} =\overline{v}_{ \uparrow } \! \left( p \right) \, .
\label{good_spinor} 
\end{alignat}

\end{appendix}

\bibliographystyle{apsrev4-2}

\begin{thebibliography}{48}%
\makeatletter
\providecommand \@ifxundefined [1]{%
 \@ifx{#1\undefined}
}%
\providecommand \@ifnum [1]{%
 \ifnum #1\expandafter \@firstoftwo
 \else \expandafter \@secondoftwo
 \fi
}%
\providecommand \@ifx [1]{%
 \ifx #1\expandafter \@firstoftwo
 \else \expandafter \@secondoftwo
 \fi
}%
\providecommand \natexlab [1]{#1}%
\providecommand \enquote  [1]{``#1''}%
\providecommand \bibnamefont  [1]{#1}%
\providecommand \bibfnamefont [1]{#1}%
\providecommand \citenamefont [1]{#1}%
\providecommand \href@noop [0]{\@secondoftwo}%
\providecommand \href [0]{\begingroup \@sanitize@url \@href}%
\providecommand \@href[1]{\@@startlink{#1}\@@href}%
\providecommand \@@href[1]{\endgroup#1\@@endlink}%
\providecommand \@sanitize@url [0]{\catcode `\\12\catcode `\$12\catcode
  `\&12\catcode `\#12\catcode `\^12\catcode `\_12\catcode `\%12\relax}%
\providecommand \@@startlink[1]{}%
\providecommand \@@endlink[0]{}%
\providecommand \url  [0]{\begingroup\@sanitize@url \@url }%
\providecommand \@url [1]{\endgroup\@href {#1}{\urlprefix }}%
\providecommand \urlprefix  [0]{URL }%
\providecommand \Eprint [0]{\href }%
\providecommand \doibase [0]{https://doi.org/}%
\providecommand \selectlanguage [0]{\@gobble}%
\providecommand \bibinfo  [0]{\@secondoftwo}%
\providecommand \bibfield  [0]{\@secondoftwo}%
\providecommand \translation [1]{[#1]}%
\providecommand \BibitemOpen [0]{}%
\providecommand \bibitemStop [0]{}%
\providecommand \bibitemNoStop [0]{.\EOS\space}%
\providecommand \EOS [0]{\spacefactor3000\relax}%
\providecommand \BibitemShut  [1]{\csname bibitem#1\endcsname}%
\let\auto@bib@innerbib\@empty
\bibitem [{\citenamefont {M\"uller}\ \emph {et~al.}(1994)\citenamefont
  {M\"uller}, \citenamefont {Robaschik}, \citenamefont {Geyer}, \citenamefont
  {Dittes},\ and\ \citenamefont {Ho\v{r}ej\v{s}i}}]{Muller:1994ses}%
  \BibitemOpen
  \bibfield  {author} {\bibinfo {author} {\bibfnamefont {D.}~\bibnamefont
  {M\"uller}}, \bibinfo {author} {\bibfnamefont {D.}~\bibnamefont {Robaschik}},
  \bibinfo {author} {\bibfnamefont {B.}~\bibnamefont {Geyer}}, \bibinfo
  {author} {\bibfnamefont {F.~M.}\ \bibnamefont {Dittes}},\ and\ \bibinfo
  {author} {\bibfnamefont {J.}~\bibnamefont {Ho\v{r}ej\v{s}i}},\ }\href
  {https://doi.org/10.1002/prop.2190420202} {\bibfield  {journal} {\bibinfo
  {journal} {Fortsch. Phys.}\ }\textbf {\bibinfo {volume} {42}},\ \bibinfo
  {pages} {101} (\bibinfo {year} {1994})},\ \Eprint
  {https://arxiv.org/abs/hep-ph/9812448} {arXiv:hep-ph/9812448} \BibitemShut
  {NoStop}%
\bibitem [{\citenamefont {Ji}(1997)}]{Ji:1996nm}%
  \BibitemOpen
  \bibfield  {author} {\bibinfo {author} {\bibfnamefont {X.-D.}\ \bibnamefont
  {Ji}},\ }\href {https://doi.org/10.1103/PhysRevD.55.7114} {\bibfield
  {journal} {\bibinfo  {journal} {Phys. Rev. D}\ }\textbf {\bibinfo {volume}
  {55}},\ \bibinfo {pages} {7114} (\bibinfo {year} {1997})},\ \Eprint
  {https://arxiv.org/abs/hep-ph/9609381} {arXiv:hep-ph/9609381} \BibitemShut
  {NoStop}%
\bibitem [{\citenamefont {Radyushkin}(1997)}]{Radyushkin:1997ki}%
  \BibitemOpen
  \bibfield  {author} {\bibinfo {author} {\bibfnamefont {A.~V.}\ \bibnamefont
  {Radyushkin}},\ }\href {https://doi.org/10.1103/PhysRevD.56.5524} {\bibfield
  {journal} {\bibinfo  {journal} {Phys. Rev. D}\ }\textbf {\bibinfo {volume}
  {56}},\ \bibinfo {pages} {5524} (\bibinfo {year} {1997})},\ \Eprint
  {https://arxiv.org/abs/hep-ph/9704207} {arXiv:hep-ph/9704207} \BibitemShut
  {NoStop}%
\bibitem [{\citenamefont {Goeke}\ \emph {et~al.}(2001)\citenamefont {Goeke},
  \citenamefont {Polyakov},\ and\ \citenamefont
  {Vanderhaeghen}}]{Goeke:2001tz}%
  \BibitemOpen
  \bibfield  {author} {\bibinfo {author} {\bibfnamefont {K.}~\bibnamefont
  {Goeke}}, \bibinfo {author} {\bibfnamefont {M.~V.}\ \bibnamefont
  {Polyakov}},\ and\ \bibinfo {author} {\bibfnamefont {M.}~\bibnamefont
  {Vanderhaeghen}},\ }\href {https://doi.org/10.1016/S0146-6410(01)00158-2}
  {\bibfield  {journal} {\bibinfo  {journal} {Prog. Part. Nucl. Phys.}\
  }\textbf {\bibinfo {volume} {47}},\ \bibinfo {pages} {401} (\bibinfo {year}
  {2001})},\ \Eprint {https://arxiv.org/abs/hep-ph/0106012}
  {arXiv:hep-ph/0106012} \BibitemShut {NoStop}%
\bibitem [{\citenamefont {Diehl}(2003)}]{Diehl_2003}%
  \BibitemOpen
  \bibfield  {author} {\bibinfo {author} {\bibfnamefont {M.}~\bibnamefont
  {Diehl}},\ }\href {https://doi.org/10.1016/j.physrep.2003.08.002} {\bibfield
  {journal} {\bibinfo  {journal} {Physics Reports}\ }\textbf {\bibinfo {volume}
  {388}},\ \bibinfo {pages} {41} (\bibinfo {year} {2003})},\ \Eprint
  {https://arxiv.org/abs/hep-ph/0307382v2} {arXiv:hep-ph/0307382v2}
  \BibitemShut {NoStop}%
\bibitem [{\citenamefont {Belitsky}\ and\ \citenamefont
  {Radyushkin}(2005)}]{Belitsky:2005qn}%
  \BibitemOpen
  \bibfield  {author} {\bibinfo {author} {\bibfnamefont {A.~V.}\ \bibnamefont
  {Belitsky}}\ and\ \bibinfo {author} {\bibfnamefont {A.~V.}\ \bibnamefont
  {Radyushkin}},\ }\href {https://doi.org/10.1016/j.physrep.2005.06.002}
  {\bibfield  {journal} {\bibinfo  {journal} {Phys. Rept.}\ }\textbf {\bibinfo
  {volume} {418}},\ \bibinfo {pages} {1} (\bibinfo {year} {2005})},\ \Eprint
  {https://arxiv.org/abs/hep-ph/0504030} {arXiv:hep-ph/0504030} \BibitemShut
  {NoStop}%
\bibitem [{\citenamefont {Boffi}\ and\ \citenamefont
  {Pasquini}(2007)}]{Boffi:2007yc}%
  \BibitemOpen
  \bibfield  {author} {\bibinfo {author} {\bibfnamefont {S.}~\bibnamefont
  {Boffi}}\ and\ \bibinfo {author} {\bibfnamefont {B.}~\bibnamefont
  {Pasquini}},\ }\href {https://doi.org/10.1393/ncr/i2007-10025-7} {\bibfield
  {journal} {\bibinfo  {journal} {Riv. Nuovo Cim.}\ }\textbf {\bibinfo {volume}
  {30}},\ \bibinfo {pages} {387} (\bibinfo {year} {2007})},\ \Eprint
  {https://arxiv.org/abs/0711.2625} {arXiv:0711.2625 [hep-ph]} \BibitemShut
  {NoStop}%
\bibitem [{\citenamefont {Berger}\ \emph {et~al.}(2002)\citenamefont {Berger},
  \citenamefont {Diehl},\ and\ \citenamefont {Pire}}]{Berger:2001xd}%
  \BibitemOpen
  \bibfield  {author} {\bibinfo {author} {\bibfnamefont {E.~R.}\ \bibnamefont
  {Berger}}, \bibinfo {author} {\bibfnamefont {M.}~\bibnamefont {Diehl}},\ and\
  \bibinfo {author} {\bibfnamefont {B.}~\bibnamefont {Pire}},\ }\href
  {https://doi.org/10.1007/s100520200917} {\bibfield  {journal} {\bibinfo
  {journal} {Eur. Phys. J. C}\ }\textbf {\bibinfo {volume} {23}},\ \bibinfo
  {pages} {675} (\bibinfo {year} {2002})},\ \Eprint
  {https://arxiv.org/abs/hep-ph/0110062} {arXiv:hep-ph/0110062} \BibitemShut
  {NoStop}%
\bibitem [{\citenamefont {Pire}\ \emph {et~al.}(2011)\citenamefont {Pire},
  \citenamefont {Szymanowski},\ and\ \citenamefont {Wagner}}]{Pire:2011st}%
  \BibitemOpen
  \bibfield  {author} {\bibinfo {author} {\bibfnamefont {B.}~\bibnamefont
  {Pire}}, \bibinfo {author} {\bibfnamefont {L.}~\bibnamefont {Szymanowski}},\
  and\ \bibinfo {author} {\bibfnamefont {J.}~\bibnamefont {Wagner}},\ }\href
  {https://doi.org/10.1103/PhysRevD.83.034009} {\bibfield  {journal} {\bibinfo
  {journal} {Phys. Rev. D}\ }\textbf {\bibinfo {volume} {83}},\ \bibinfo
  {pages} {034009} (\bibinfo {year} {2011})},\ \Eprint
  {https://arxiv.org/abs/1101.0555} {arXiv:1101.0555 [hep-ph]} \BibitemShut
  {NoStop}%
\bibitem [{\citenamefont {Mueller}\ \emph {et~al.}(2012)\citenamefont
  {Mueller}, \citenamefont {Pire}, \citenamefont {Szymanowski},\ and\
  \citenamefont {Wagner}}]{Mueller:2012sma}%
  \BibitemOpen
  \bibfield  {author} {\bibinfo {author} {\bibfnamefont {D.}~\bibnamefont
  {Mueller}}, \bibinfo {author} {\bibfnamefont {B.}~\bibnamefont {Pire}},
  \bibinfo {author} {\bibfnamefont {L.}~\bibnamefont {Szymanowski}},\ and\
  \bibinfo {author} {\bibfnamefont {J.}~\bibnamefont {Wagner}},\ }\href
  {https://doi.org/10.1103/PhysRevD.86.031502} {\bibfield  {journal} {\bibinfo
  {journal} {Phys. Rev. D}\ }\textbf {\bibinfo {volume} {86}},\ \bibinfo
  {pages} {031502} (\bibinfo {year} {2012})},\ \Eprint
  {https://arxiv.org/abs/1203.4392} {arXiv:1203.4392 [hep-ph]} \BibitemShut
  {NoStop}%
\bibitem [{\citenamefont {Moutarde}\ \emph {et~al.}(2013)\citenamefont
  {Moutarde}, \citenamefont {Pire}, \citenamefont {Sabatie}, \citenamefont
  {Szymanowski},\ and\ \citenamefont {Wagner}}]{Moutarde:2013qs}%
  \BibitemOpen
  \bibfield  {author} {\bibinfo {author} {\bibfnamefont {H.}~\bibnamefont
  {Moutarde}}, \bibinfo {author} {\bibfnamefont {B.}~\bibnamefont {Pire}},
  \bibinfo {author} {\bibfnamefont {F.}~\bibnamefont {Sabatie}}, \bibinfo
  {author} {\bibfnamefont {L.}~\bibnamefont {Szymanowski}},\ and\ \bibinfo
  {author} {\bibfnamefont {J.}~\bibnamefont {Wagner}},\ }\href
  {https://doi.org/10.1103/PhysRevD.87.054029} {\bibfield  {journal} {\bibinfo
  {journal} {Phys. Rev. D}\ }\textbf {\bibinfo {volume} {87}},\ \bibinfo
  {pages} {054029} (\bibinfo {year} {2013})},\ \Eprint
  {https://arxiv.org/abs/1301.3819} {arXiv:1301.3819 [hep-ph]} \BibitemShut
  {NoStop}%
\bibitem [{\citenamefont {Pire}\ \emph {et~al.}(2021)\citenamefont {Pire},
  \citenamefont {Semenov-Tian-Shansky},\ and\ \citenamefont
  {Szymanowski}}]{Pire:2021hbl}%
  \BibitemOpen
  \bibfield  {author} {\bibinfo {author} {\bibfnamefont {B.}~\bibnamefont
  {Pire}}, \bibinfo {author} {\bibfnamefont {K.}~\bibnamefont
  {Semenov-Tian-Shansky}},\ and\ \bibinfo {author} {\bibfnamefont
  {L.}~\bibnamefont {Szymanowski}},\ }\href
  {https://doi.org/10.1016/j.physrep.2021.09.002} {\bibfield  {journal}
  {\bibinfo  {journal} {Phys. Rept.}\ }\textbf {\bibinfo {volume} {940}},\
  \bibinfo {pages} {1} (\bibinfo {year} {2021})},\ \Eprint
  {https://arxiv.org/abs/2103.01079} {arXiv:2103.01079 [hep-ph]} \BibitemShut
  {NoStop}%
\bibitem [{\citenamefont {Pire}\ \emph {et~al.}(2022)\citenamefont {Pire},
  \citenamefont {Semenov-Tian-Shansky}, \citenamefont {Shaikhutdinova},\ and\
  \citenamefont {Szymanowski}}]{Pire:2022fbi}%
  \BibitemOpen
  \bibfield  {author} {\bibinfo {author} {\bibfnamefont {B.}~\bibnamefont
  {Pire}}, \bibinfo {author} {\bibfnamefont {K.~M.}\ \bibnamefont
  {Semenov-Tian-Shansky}}, \bibinfo {author} {\bibfnamefont {A.~A.}\
  \bibnamefont {Shaikhutdinova}},\ and\ \bibinfo {author} {\bibfnamefont
  {L.}~\bibnamefont {Szymanowski}},\ }\href
  {https://doi.org/10.1140/epjc/s10052-022-10587-4} {\bibfield  {journal}
  {\bibinfo  {journal} {Eur. Phys. J. C}\ }\textbf {\bibinfo {volume} {82}},\
  \bibinfo {pages} {656} (\bibinfo {year} {2022})},\ \Eprint
  {https://arxiv.org/abs/2201.12853} {arXiv:2201.12853 [hep-ph]} \BibitemShut
  {NoStop}%
\bibitem [{\citenamefont {Pire}\ \emph {et~al.}(2023)\citenamefont {Pire},
  \citenamefont {Semenov-Tian-Shansky}, \citenamefont {Shaikhutdinova},\ and\
  \citenamefont {Szymanowski}}]{Pire:2022kwu}%
  \BibitemOpen
  \bibfield  {author} {\bibinfo {author} {\bibfnamefont {B.}~\bibnamefont
  {Pire}}, \bibinfo {author} {\bibfnamefont {K.~M.}\ \bibnamefont
  {Semenov-Tian-Shansky}}, \bibinfo {author} {\bibfnamefont {A.~A.}\
  \bibnamefont {Shaikhutdinova}},\ and\ \bibinfo {author} {\bibfnamefont
  {L.}~\bibnamefont {Szymanowski}},\ }\href
  {https://doi.org/10.1007/s43673-023-00094-3} {\bibfield  {journal} {\bibinfo
  {journal} {AAPPS Bull.}\ }\textbf {\bibinfo {volume} {33}},\ \bibinfo {pages}
  {26} (\bibinfo {year} {2023})},\ \Eprint {https://arxiv.org/abs/2212.07688}
  {arXiv:2212.07688 [hep-ph]} \BibitemShut {NoStop}%
\bibitem [{\citenamefont {Lansberg}\ \emph
  {et~al.}(2012{\natexlab{a}})\citenamefont {Lansberg}, \citenamefont {Pire},
  \citenamefont {Semenov-Tian-Shansky},\ and\ \citenamefont
  {Szymanowski}}]{Lansberg:2011aa}%
  \BibitemOpen
  \bibfield  {author} {\bibinfo {author} {\bibfnamefont {J.~P.}\ \bibnamefont
  {Lansberg}}, \bibinfo {author} {\bibfnamefont {B.}~\bibnamefont {Pire}},
  \bibinfo {author} {\bibfnamefont {K.}~\bibnamefont {Semenov-Tian-Shansky}},\
  and\ \bibinfo {author} {\bibfnamefont {L.}~\bibnamefont {Szymanowski}},\
  }\href {https://doi.org/10.1103/PhysRevD.85.054021} {\bibfield  {journal}
  {\bibinfo  {journal} {Phys. Rev. D}\ }\textbf {\bibinfo {volume} {85}},\
  \bibinfo {pages} {054021} (\bibinfo {year} {2012}{\natexlab{a}})},\ \Eprint
  {https://arxiv.org/abs/1112.3570} {arXiv:1112.3570 [hep-ph]} \BibitemShut
  {NoStop}%
\bibitem [{\citenamefont {Lansberg}\ \emph
  {et~al.}(2012{\natexlab{b}})\citenamefont {Lansberg}, \citenamefont {Pire},
  \citenamefont {Semenov-Tian-Shansky},\ and\ \citenamefont
  {Szymanowski}}]{Lansberg:2012ha}%
  \BibitemOpen
  \bibfield  {author} {\bibinfo {author} {\bibfnamefont {J.~P.}\ \bibnamefont
  {Lansberg}}, \bibinfo {author} {\bibfnamefont {B.}~\bibnamefont {Pire}},
  \bibinfo {author} {\bibfnamefont {K.}~\bibnamefont {Semenov-Tian-Shansky}},\
  and\ \bibinfo {author} {\bibfnamefont {L.}~\bibnamefont {Szymanowski}},\
  }\href {https://doi.org/10.1103/PhysRevD.86.114033} {\bibfield  {journal}
  {\bibinfo  {journal} {Phys. Rev. D}\ }\textbf {\bibinfo {volume} {86}},\
  \bibinfo {pages} {114033} (\bibinfo {year} {2012}{\natexlab{b}})},\ \bibinfo
  {note} {[Erratum: Phys.Rev.D 87, 059902 (2013)]},\ \Eprint
  {https://arxiv.org/abs/1210.0126} {arXiv:1210.0126 [hep-ph]} \BibitemShut
  {NoStop}%
\bibitem [{\citenamefont {Pire}\ \emph {et~al.}(2013)\citenamefont {Pire},
  \citenamefont {Semenov-Tian-Shansky},\ and\ \citenamefont
  {Szymanowski}}]{Pire:2013jva}%
  \BibitemOpen
  \bibfield  {author} {\bibinfo {author} {\bibfnamefont {B.}~\bibnamefont
  {Pire}}, \bibinfo {author} {\bibfnamefont {K.}~\bibnamefont
  {Semenov-Tian-Shansky}},\ and\ \bibinfo {author} {\bibfnamefont
  {L.}~\bibnamefont {Szymanowski}},\ }\href
  {https://doi.org/10.1016/j.physletb.2013.06.015} {\bibfield  {journal}
  {\bibinfo  {journal} {Phys. Lett. B}\ }\textbf {\bibinfo {volume} {724}},\
  \bibinfo {pages} {99} (\bibinfo {year} {2013})},\ \bibinfo {note} {[Erratum:
  Phys.Lett.B 764, 335--335 (2017)]},\ \Eprint
  {https://arxiv.org/abs/1304.6298} {arXiv:1304.6298 [hep-ph]} \BibitemShut
  {NoStop}%
\bibitem [{\citenamefont {Pire}\ \emph {et~al.}(2015)\citenamefont {Pire},
  \citenamefont {Semenov-Tian-Shansky},\ and\ \citenamefont
  {Szymanowski}}]{Pire:2015kxa}%
  \BibitemOpen
  \bibfield  {author} {\bibinfo {author} {\bibfnamefont {B.}~\bibnamefont
  {Pire}}, \bibinfo {author} {\bibfnamefont {K.}~\bibnamefont
  {Semenov-Tian-Shansky}},\ and\ \bibinfo {author} {\bibfnamefont
  {L.}~\bibnamefont {Szymanowski}},\ }\href
  {https://doi.org/10.1103/PhysRevD.91.094006} {\bibfield  {journal} {\bibinfo
  {journal} {Phys. Rev. D}\ }\textbf {\bibinfo {volume} {91}},\ \bibinfo
  {pages} {094006} (\bibinfo {year} {2015})},\ \bibinfo {note} {[Erratum:
  Phys.Rev.D 106, 099901 (2022)]},\ \Eprint {https://arxiv.org/abs/1503.02012}
  {arXiv:1503.02012 [hep-ph]} \BibitemShut {NoStop}%
\bibitem [{\citenamefont {Pasquini}\ \emph {et~al.}(2009)\citenamefont
  {Pasquini}, \citenamefont {Pincetti},\ and\ \citenamefont
  {Boffi}}]{Pasquini:2009ki}%
  \BibitemOpen
  \bibfield  {author} {\bibinfo {author} {\bibfnamefont {B.}~\bibnamefont
  {Pasquini}}, \bibinfo {author} {\bibfnamefont {M.}~\bibnamefont {Pincetti}},\
  and\ \bibinfo {author} {\bibfnamefont {S.}~\bibnamefont {Boffi}},\ }\href
  {https://doi.org/10.1103/PhysRevD.80.014017} {\bibfield  {journal} {\bibinfo
  {journal} {Phys. Rev. D}\ }\textbf {\bibinfo {volume} {80}},\ \bibinfo
  {pages} {014017} (\bibinfo {year} {2009})},\ \Eprint
  {https://arxiv.org/abs/0905.4018} {arXiv:0905.4018 [hep-ph]} \BibitemShut
  {NoStop}%
\bibitem [{\citenamefont {Chatagnon}\ \emph {et~al.}(2021)\citenamefont
  {Chatagnon} \emph {et~al.}}]{CLAS:2021lky}%
  \BibitemOpen
  \bibfield  {author} {\bibinfo {author} {\bibfnamefont {P.}~\bibnamefont
  {Chatagnon}} \emph {et~al.} (\bibinfo {collaboration} {CLAS}),\ }\href
  {https://doi.org/10.1103/PhysRevLett.127.262501} {\bibfield  {journal}
  {\bibinfo  {journal} {Phys. Rev. Lett.}\ }\textbf {\bibinfo {volume} {127}},\
  \bibinfo {pages} {262501} (\bibinfo {year} {2021})},\ \Eprint
  {https://arxiv.org/abs/2108.11746} {arXiv:2108.11746 [hep-ex]} \BibitemShut
  {NoStop}%
\bibitem [{\citenamefont {Dirac}(1949)}]{Dirac:1949cp}%
  \BibitemOpen
  \bibfield  {author} {\bibinfo {author} {\bibfnamefont {P.~A.~M.}\
  \bibnamefont {Dirac}},\ }\href {https://doi.org/10.1103/RevModPhys.21.392}
  {\bibfield  {journal} {\bibinfo  {journal} {Rev. Mod. Phys.}\ }\textbf
  {\bibinfo {volume} {21}},\ \bibinfo {pages} {392} (\bibinfo {year}
  {1949})}\BibitemShut {NoStop}%
\bibitem [{\citenamefont {Brodsky}\ and\ \citenamefont
  {Lepage}(1989)}]{Brodsky:1989pv}%
  \BibitemOpen
  \bibfield  {author} {\bibinfo {author} {\bibfnamefont {S.~J.}\ \bibnamefont
  {Brodsky}}\ and\ \bibinfo {author} {\bibfnamefont {G.~P.}\ \bibnamefont
  {Lepage}},\ }\href {https://doi.org/10.1142/9789814503266_0002} {\bibfield
  {journal} {\bibinfo  {journal} {Adv. Ser. Direct. High Energy Phys.}\
  }\textbf {\bibinfo {volume} {5}},\ \bibinfo {pages} {93} (\bibinfo {year}
  {1989})}\BibitemShut {NoStop}%
\bibitem [{\citenamefont {Brodsky}\ \emph {et~al.}(1998)\citenamefont
  {Brodsky}, \citenamefont {Pauli},\ and\ \citenamefont
  {Pinsky}}]{Brodsky:1997de}%
  \BibitemOpen
  \bibfield  {author} {\bibinfo {author} {\bibfnamefont {S.~J.}\ \bibnamefont
  {Brodsky}}, \bibinfo {author} {\bibfnamefont {H.-C.}\ \bibnamefont {Pauli}},\
  and\ \bibinfo {author} {\bibfnamefont {S.~S.}\ \bibnamefont {Pinsky}},\
  }\href {https://doi.org/10.1016/S0370-1573(97)00089-6} {\bibfield  {journal}
  {\bibinfo  {journal} {Phys. Rept.}\ }\textbf {\bibinfo {volume} {301}},\
  \bibinfo {pages} {299} (\bibinfo {year} {1998})},\ \Eprint
  {https://arxiv.org/abs/hep-ph/9705477} {arXiv:hep-ph/9705477} \BibitemShut
  {NoStop}%
\bibitem [{\citenamefont {Kovchegov}\ and\ \citenamefont
  {Levin}(2013)}]{Kovchegov:2012mbw}%
  \BibitemOpen
  \bibfield  {author} {\bibinfo {author} {\bibfnamefont {Y.~V.}\ \bibnamefont
  {Kovchegov}}\ and\ \bibinfo {author} {\bibfnamefont {E.}~\bibnamefont
  {Levin}},\ }\href {https://doi.org/10.1017/9781009291446} {\emph {\bibinfo
  {title} {{Quantum Chromodynamics at High Energy}}}},\ Vol.~\bibinfo {volume}
  {33}\ (\bibinfo  {publisher} {Oxford University Press},\ \bibinfo {year}
  {2013})\BibitemShut {NoStop}%
\bibitem [{\citenamefont {Boffi}\ \emph {et~al.}(2003)\citenamefont {Boffi},
  \citenamefont {Pasquini},\ and\ \citenamefont {Traini}}]{Boffi:2002yy}%
  \BibitemOpen
  \bibfield  {author} {\bibinfo {author} {\bibfnamefont {S.}~\bibnamefont
  {Boffi}}, \bibinfo {author} {\bibfnamefont {B.}~\bibnamefont {Pasquini}},\
  and\ \bibinfo {author} {\bibfnamefont {M.}~\bibnamefont {Traini}},\ }\href
  {https://doi.org/10.1016/S0550-3213(02)01016-7} {\bibfield  {journal}
  {\bibinfo  {journal} {Nucl. Phys. B}\ }\textbf {\bibinfo {volume} {649}},\
  \bibinfo {pages} {243} (\bibinfo {year} {2003})},\ \Eprint
  {https://arxiv.org/abs/hep-ph/0207340} {arXiv:hep-ph/0207340} \BibitemShut
  {NoStop}%
\bibitem [{\citenamefont {Boffi}\ \emph {et~al.}(2004)\citenamefont {Boffi},
  \citenamefont {Pasquini},\ and\ \citenamefont {Traini}}]{Boffi:2003yj}%
  \BibitemOpen
  \bibfield  {author} {\bibinfo {author} {\bibfnamefont {S.}~\bibnamefont
  {Boffi}}, \bibinfo {author} {\bibfnamefont {B.}~\bibnamefont {Pasquini}},\
  and\ \bibinfo {author} {\bibfnamefont {M.}~\bibnamefont {Traini}},\ }\href
  {https://doi.org/10.1016/j.nuclphysb.2003.12.037} {\bibfield  {journal}
  {\bibinfo  {journal} {Nucl. Phys. B}\ }\textbf {\bibinfo {volume} {680}},\
  \bibinfo {pages} {147} (\bibinfo {year} {2004})},\ \Eprint
  {https://arxiv.org/abs/hep-ph/0311016} {arXiv:hep-ph/0311016} \BibitemShut
  {NoStop}%
\bibitem [{\citenamefont {Pasquini}\ \emph
  {et~al.}(2005{\natexlab{a}})\citenamefont {Pasquini}, \citenamefont
  {Traini},\ and\ \citenamefont {Boffi}}]{Pasquini:2004gc}%
  \BibitemOpen
  \bibfield  {author} {\bibinfo {author} {\bibfnamefont {B.}~\bibnamefont
  {Pasquini}}, \bibinfo {author} {\bibfnamefont {M.}~\bibnamefont {Traini}},\
  and\ \bibinfo {author} {\bibfnamefont {S.}~\bibnamefont {Boffi}},\ }\href
  {https://doi.org/10.1103/PhysRevD.71.034022} {\bibfield  {journal} {\bibinfo
  {journal} {Phys. Rev. D}\ }\textbf {\bibinfo {volume} {71}},\ \bibinfo
  {pages} {034022} (\bibinfo {year} {2005}{\natexlab{a}})},\ \Eprint
  {https://arxiv.org/abs/hep-ph/0407228} {arXiv:hep-ph/0407228} \BibitemShut
  {NoStop}%
\bibitem [{\citenamefont {Pasquini}\ \emph
  {et~al.}(2005{\natexlab{b}})\citenamefont {Pasquini}, \citenamefont
  {Pincetti},\ and\ \citenamefont {Boffi}}]{Pasquini:2005dk}%
  \BibitemOpen
  \bibfield  {author} {\bibinfo {author} {\bibfnamefont {B.}~\bibnamefont
  {Pasquini}}, \bibinfo {author} {\bibfnamefont {M.}~\bibnamefont {Pincetti}},\
  and\ \bibinfo {author} {\bibfnamefont {S.}~\bibnamefont {Boffi}},\ }\href
  {https://doi.org/10.1103/PhysRevD.72.094029} {\bibfield  {journal} {\bibinfo
  {journal} {Phys. Rev. D}\ }\textbf {\bibinfo {volume} {72}},\ \bibinfo
  {pages} {094029} (\bibinfo {year} {2005}{\natexlab{b}})},\ \Eprint
  {https://arxiv.org/abs/hep-ph/0510376} {arXiv:hep-ph/0510376} \BibitemShut
  {NoStop}%
\bibitem [{\citenamefont {Pasquini}\ and\ \citenamefont
  {Boffi}(2006)}]{Pasquini:2006dv}%
  \BibitemOpen
  \bibfield  {author} {\bibinfo {author} {\bibfnamefont {B.}~\bibnamefont
  {Pasquini}}\ and\ \bibinfo {author} {\bibfnamefont {S.}~\bibnamefont
  {Boffi}},\ }\href {https://doi.org/10.1103/PhysRevD.73.094001} {\bibfield
  {journal} {\bibinfo  {journal} {Phys. Rev. D}\ }\textbf {\bibinfo {volume}
  {73}},\ \bibinfo {pages} {094001} (\bibinfo {year} {2006})},\ \Eprint
  {https://arxiv.org/abs/hep-ph/0601177} {arXiv:hep-ph/0601177} \BibitemShut
  {NoStop}%
\bibitem [{\citenamefont {Pasquini}\ and\ \citenamefont
  {Boffi}(2007)}]{Pasquini:2007xz}%
  \BibitemOpen
  \bibfield  {author} {\bibinfo {author} {\bibfnamefont {B.}~\bibnamefont
  {Pasquini}}\ and\ \bibinfo {author} {\bibfnamefont {S.}~\bibnamefont
  {Boffi}},\ }\href {https://doi.org/10.1016/j.physletb.2007.07.037} {\bibfield
   {journal} {\bibinfo  {journal} {Phys. Lett. B}\ }\textbf {\bibinfo {volume}
  {653}},\ \bibinfo {pages} {23} (\bibinfo {year} {2007})},\ \Eprint
  {https://arxiv.org/abs/0705.4345} {arXiv:0705.4345 [hep-ph]} \BibitemShut
  {NoStop}%
\bibitem [{\citenamefont {Gribov}\ and\ \citenamefont
  {Lipatov}(1972)}]{Gribov:1972ri}%
  \BibitemOpen
  \bibfield  {author} {\bibinfo {author} {\bibfnamefont {V.~N.}\ \bibnamefont
  {Gribov}}\ and\ \bibinfo {author} {\bibfnamefont {L.~N.}\ \bibnamefont
  {Lipatov}},\ }\href@noop {} {\bibfield  {journal} {\bibinfo  {journal} {Sov.
  J. Nucl. Phys.}\ }\textbf {\bibinfo {volume} {15}},\ \bibinfo {pages} {438}
  (\bibinfo {year} {1972})}\BibitemShut {NoStop}%
\bibitem [{\citenamefont {Altarelli}\ and\ \citenamefont
  {Parisi}(1977)}]{Altarelli:1977zs}%
  \BibitemOpen
  \bibfield  {author} {\bibinfo {author} {\bibfnamefont {G.}~\bibnamefont
  {Altarelli}}\ and\ \bibinfo {author} {\bibfnamefont {G.}~\bibnamefont
  {Parisi}},\ }\href {https://doi.org/10.1016/0550-3213(77)90384-4} {\bibfield
  {journal} {\bibinfo  {journal} {Nucl. Phys. B}\ }\textbf {\bibinfo {volume}
  {126}},\ \bibinfo {pages} {298} (\bibinfo {year} {1977})}\BibitemShut
  {NoStop}%
\bibitem [{\citenamefont {Dokshitzer}(1977)}]{Dokshitzer:1977sg}%
  \BibitemOpen
  \bibfield  {author} {\bibinfo {author} {\bibfnamefont {Y.~L.}\ \bibnamefont
  {Dokshitzer}},\ }\href@noop {} {\bibfield  {journal} {\bibinfo  {journal}
  {Sov. Phys. JETP}\ }\textbf {\bibinfo {volume} {46}},\ \bibinfo {pages} {641}
  (\bibinfo {year} {1977})}\BibitemShut {NoStop}%
\bibitem [{\citenamefont {Efremov}\ and\ \citenamefont
  {Radyushkin}(1980{\natexlab{a}})}]{Efremov:1978rn}%
  \BibitemOpen
  \bibfield  {author} {\bibinfo {author} {\bibfnamefont {A.~V.}\ \bibnamefont
  {Efremov}}\ and\ \bibinfo {author} {\bibfnamefont {A.~V.}\ \bibnamefont
  {Radyushkin}},\ }\href {https://doi.org/10.1007/BF01032111} {\bibfield
  {journal} {\bibinfo  {journal} {Theor. Math. Phys.}\ }\textbf {\bibinfo
  {volume} {42}},\ \bibinfo {pages} {97} (\bibinfo {year}
  {1980}{\natexlab{a}})}\BibitemShut {NoStop}%
\bibitem [{\citenamefont {Efremov}\ and\ \citenamefont
  {Radyushkin}(1980{\natexlab{b}})}]{Efremov:1979qk}%
  \BibitemOpen
  \bibfield  {author} {\bibinfo {author} {\bibfnamefont {A.~V.}\ \bibnamefont
  {Efremov}}\ and\ \bibinfo {author} {\bibfnamefont {A.~V.}\ \bibnamefont
  {Radyushkin}},\ }\href {https://doi.org/10.1016/0370-2693(80)90869-2}
  {\bibfield  {journal} {\bibinfo  {journal} {Phys. Lett. B}\ }\textbf
  {\bibinfo {volume} {94}},\ \bibinfo {pages} {245} (\bibinfo {year}
  {1980}{\natexlab{b}})}\BibitemShut {NoStop}%
\bibitem [{\citenamefont {Lepage}\ and\ \citenamefont
  {Brodsky}(1979{\natexlab{a}})}]{Lepage:1979zb}%
  \BibitemOpen
  \bibfield  {author} {\bibinfo {author} {\bibfnamefont {G.~P.}\ \bibnamefont
  {Lepage}}\ and\ \bibinfo {author} {\bibfnamefont {S.~J.}\ \bibnamefont
  {Brodsky}},\ }\href {https://doi.org/10.1016/0370-2693(79)90554-9} {\bibfield
   {journal} {\bibinfo  {journal} {Phys. Lett. B}\ }\textbf {\bibinfo {volume}
  {87}},\ \bibinfo {pages} {359} (\bibinfo {year}
  {1979}{\natexlab{a}})}\BibitemShut {NoStop}%
\bibitem [{\citenamefont {Lepage}\ and\ \citenamefont
  {Brodsky}(1979{\natexlab{b}})}]{Lepage:1979za}%
  \BibitemOpen
  \bibfield  {author} {\bibinfo {author} {\bibfnamefont {G.~P.}\ \bibnamefont
  {Lepage}}\ and\ \bibinfo {author} {\bibfnamefont {S.~J.}\ \bibnamefont
  {Brodsky}},\ }\href {https://doi.org/10.1103/PhysRevLett.43.1625.2}
  {\bibfield  {journal} {\bibinfo  {journal} {Phys. Rev. Lett.}\ }\textbf
  {\bibinfo {volume} {43}},\ \bibinfo {pages} {545} (\bibinfo {year}
  {1979}{\natexlab{b}})},\ \bibinfo {note} {[Erratum: Phys.Rev.Lett. 43,
  1625--1626 (1979)]}\BibitemShut {NoStop}%
\bibitem [{\citenamefont {Lepage}\ and\ \citenamefont
  {Brodsky}(1980)}]{Lepage:1980fj}%
  \BibitemOpen
  \bibfield  {author} {\bibinfo {author} {\bibfnamefont {G.~P.}\ \bibnamefont
  {Lepage}}\ and\ \bibinfo {author} {\bibfnamefont {S.~J.}\ \bibnamefont
  {Brodsky}},\ }\href {https://doi.org/10.1103/PhysRevD.22.2157} {\bibfield
  {journal} {\bibinfo  {journal} {Phys. Rev. D}\ }\textbf {\bibinfo {volume}
  {22}},\ \bibinfo {pages} {2157} (\bibinfo {year} {1980})}\BibitemShut
  {NoStop}%
\bibitem [{\citenamefont {Keister}\ and\ \citenamefont
  {Polyzou}(1991)}]{Keister:1991sb}%
  \BibitemOpen
  \bibfield  {author} {\bibinfo {author} {\bibfnamefont {B.~D.}\ \bibnamefont
  {Keister}}\ and\ \bibinfo {author} {\bibfnamefont {W.~N.}\ \bibnamefont
  {Polyzou}},\ }\href@noop {} {\bibfield  {journal} {\bibinfo  {journal} {Adv.
  Nucl. Phys.}\ }\textbf {\bibinfo {volume} {20}},\ \bibinfo {pages} {225}
  (\bibinfo {year} {1991})}\BibitemShut {NoStop}%
\bibitem [{\citenamefont {Pasquini}\ \emph {et~al.}(2008)\citenamefont
  {Pasquini}, \citenamefont {Cazzaniga},\ and\ \citenamefont
  {Boffi}}]{Pasquini:2008ax}%
  \BibitemOpen
  \bibfield  {author} {\bibinfo {author} {\bibfnamefont {B.}~\bibnamefont
  {Pasquini}}, \bibinfo {author} {\bibfnamefont {S.}~\bibnamefont
  {Cazzaniga}},\ and\ \bibinfo {author} {\bibfnamefont {S.}~\bibnamefont
  {Boffi}},\ }\href {https://doi.org/10.1103/PhysRevD.78.034025} {\bibfield
  {journal} {\bibinfo  {journal} {Phys. Rev. D}\ }\textbf {\bibinfo {volume}
  {78}},\ \bibinfo {pages} {034025} (\bibinfo {year} {2008})},\ \Eprint
  {https://arxiv.org/abs/0806.2298} {arXiv:0806.2298 [hep-ph]} \BibitemShut
  {NoStop}%
\bibitem [{\citenamefont {Pasquini}\ and\ \citenamefont
  {Yuan}(2010)}]{Pasquini:2010af}%
  \BibitemOpen
  \bibfield  {author} {\bibinfo {author} {\bibfnamefont {B.}~\bibnamefont
  {Pasquini}}\ and\ \bibinfo {author} {\bibfnamefont {F.}~\bibnamefont
  {Yuan}},\ }\href {https://doi.org/10.1103/PhysRevD.81.114013} {\bibfield
  {journal} {\bibinfo  {journal} {Phys. Rev. D}\ }\textbf {\bibinfo {volume}
  {81}},\ \bibinfo {pages} {114013} (\bibinfo {year} {2010})},\ \Eprint
  {https://arxiv.org/abs/1001.5398} {arXiv:1001.5398 [hep-ph]} \BibitemShut
  {NoStop}%
\bibitem [{\citenamefont {Boffi}\ \emph {et~al.}(2009)\citenamefont {Boffi},
  \citenamefont {Efremov}, \citenamefont {Pasquini},\ and\ \citenamefont
  {Schweitzer}}]{Boffi:2009sh}%
  \BibitemOpen
  \bibfield  {author} {\bibinfo {author} {\bibfnamefont {S.}~\bibnamefont
  {Boffi}}, \bibinfo {author} {\bibfnamefont {A.~V.}\ \bibnamefont {Efremov}},
  \bibinfo {author} {\bibfnamefont {B.}~\bibnamefont {Pasquini}},\ and\
  \bibinfo {author} {\bibfnamefont {P.}~\bibnamefont {Schweitzer}},\ }\href
  {https://doi.org/10.1103/PhysRevD.79.094012} {\bibfield  {journal} {\bibinfo
  {journal} {Phys. Rev. D}\ }\textbf {\bibinfo {volume} {79}},\ \bibinfo
  {pages} {094012} (\bibinfo {year} {2009})},\ \Eprint
  {https://arxiv.org/abs/0903.1271} {arXiv:0903.1271 [hep-ph]} \BibitemShut
  {NoStop}%
\bibitem [{\citenamefont {Pasquini}\ and\ \citenamefont
  {Schweitzer}(2011)}]{Pasquini:2011tk}%
  \BibitemOpen
  \bibfield  {author} {\bibinfo {author} {\bibfnamefont {B.}~\bibnamefont
  {Pasquini}}\ and\ \bibinfo {author} {\bibfnamefont {P.}~\bibnamefont
  {Schweitzer}},\ }\href {https://doi.org/10.1103/PhysRevD.83.114044}
  {\bibfield  {journal} {\bibinfo  {journal} {Phys. Rev. D}\ }\textbf {\bibinfo
  {volume} {83}},\ \bibinfo {pages} {114044} (\bibinfo {year} {2011})},\
  \Eprint {https://arxiv.org/abs/1103.5977} {arXiv:1103.5977 [hep-ph]}
  \BibitemShut {NoStop}%
\bibitem [{\citenamefont {Schlumpf}(1994)}]{Schlumpf:1992pp}%
  \BibitemOpen
  \bibfield  {author} {\bibinfo {author} {\bibfnamefont {F.}~\bibnamefont
  {Schlumpf}},\ }\href {https://doi.org/10.1088/0954-3899/20/1/024} {\bibfield
  {journal} {\bibinfo  {journal} {J. Phys. G}\ }\textbf {\bibinfo {volume}
  {20}},\ \bibinfo {pages} {237} (\bibinfo {year} {1994})},\ \Eprint
  {https://arxiv.org/abs/hep-ph/9301233} {arXiv:hep-ph/9301233} \BibitemShut
  {NoStop}%
\bibitem [{\citenamefont {Park}\ \emph {et~al.}(2018)\citenamefont {Park} \emph
  {et~al.}}]{CLAS:2017rgp}%
  \BibitemOpen
  \bibfield  {author} {\bibinfo {author} {\bibfnamefont {K.}~\bibnamefont
  {Park}} \emph {et~al.} (\bibinfo {collaboration} {CLAS}),\ }\href
  {https://doi.org/10.1016/j.physletb.2018.03.026} {\bibfield  {journal}
  {\bibinfo  {journal} {Phys. Lett. B}\ }\textbf {\bibinfo {volume} {780}},\
  \bibinfo {pages} {340} (\bibinfo {year} {2018})},\ \Eprint
  {https://arxiv.org/abs/1711.08486} {arXiv:1711.08486 [nucl-ex]} \BibitemShut
  {NoStop}%
\bibitem [{\citenamefont {Li}\ \emph {et~al.}(2019)\citenamefont {Li} \emph
  {et~al.}}]{JeffersonLabFp:2019gpp}%
  \BibitemOpen
  \bibfield  {author} {\bibinfo {author} {\bibfnamefont {W.~B.}\ \bibnamefont
  {Li}} \emph {et~al.} (\bibinfo {collaboration} {Jefferson Lab
  F\ensuremath{\pi}}),\ }\href {https://doi.org/10.1103/PhysRevLett.123.182501}
  {\bibfield  {journal} {\bibinfo  {journal} {Phys. Rev. Lett.}\ }\textbf
  {\bibinfo {volume} {123}},\ \bibinfo {pages} {182501} (\bibinfo {year}
  {2019})},\ \Eprint {https://arxiv.org/abs/1910.00464} {arXiv:1910.00464
  [nucl-ex]} \BibitemShut {NoStop}%
\bibitem [{\citenamefont {Diehl}\ \emph {et~al.}(2020)\citenamefont {Diehl}
  \emph {et~al.}}]{CLAS:2020yqf}%
  \BibitemOpen
  \bibfield  {author} {\bibinfo {author} {\bibfnamefont {S.}~\bibnamefont
  {Diehl}} \emph {et~al.} (\bibinfo {collaboration} {CLAS}),\ }\href
  {https://doi.org/10.1103/PhysRevLett.125.182001} {\bibfield  {journal}
  {\bibinfo  {journal} {Phys. Rev. Lett.}\ }\textbf {\bibinfo {volume} {125}},\
  \bibinfo {pages} {182001} (\bibinfo {year} {2020})},\ \Eprint
  {https://arxiv.org/abs/2007.15677} {arXiv:2007.15677 [nucl-ex]} \BibitemShut
  {NoStop}%
\bibitem [{\citenamefont {Abdul~Khalek}\ \emph {et~al.}(2022)\citenamefont
  {Abdul~Khalek} \emph {et~al.}}]{AbdulKhalek:2021gbh}%
  \BibitemOpen
  \bibfield  {author} {\bibinfo {author} {\bibfnamefont {R.}~\bibnamefont
  {Abdul~Khalek}} \emph {et~al.},\ }\href
  {https://doi.org/10.1016/j.nuclphysa.2022.122447} {\bibfield  {journal}
  {\bibinfo  {journal} {Nucl. Phys. A}\ }\textbf {\bibinfo {volume} {1026}},\
  \bibinfo {pages} {122447} (\bibinfo {year} {2022})},\ \Eprint
  {https://arxiv.org/abs/2103.05419} {arXiv:2103.05419 [physics.ins-det]}
  \BibitemShut {NoStop}%
\end{thebibliography}
%

\end{document}